\documentclass[pdflatex,sn-mathphys-num]{sn-jnl}

\usepackage{graphicx}
\usepackage{setspace}
\usepackage{lineno}
\usepackage{multirow}
\usepackage{amsmath,amssymb,amsfonts}
\usepackage{amsthm}
\usepackage{mathrsfs}
\usepackage[title]{appendix}
\usepackage{xcolor}
\usepackage{float}
\usepackage{textcomp}
\usepackage{manyfoot}
\usepackage{booktabs}
\usepackage{algorithm}
\usepackage{algorithmicx}
\usepackage{algpseudocode}
\usepackage{listings}
\usepackage{siunitx}

\theoremstyle{thmstyleone}

\theoremstyle{thmstyletwo}

\theoremstyle{thmstylethree}

\begin{document}

\title[Hyperlocal urban NO$_2$ hotspot modeling driven by microscopic traffic data]{Hyperlocal urban NO$_2$ hotspot modeling driven by microscopic traffic data}

\author*[1,2,3]{\fnm{Michael} \sur{Weger}}\email{michael.weger@ufz.de}
\author[1,2]{\fnm{Thomas} \sur{Trabert}}\email{thomas.trabert@ufz.de}
\author[1,2]{\fnm{Timo} \sur{Houben}}\email{timo.houben@ufz.de}
\author[4]{\fnm{Alexander} \sur{Sohr}}\email{alexander.sohr@dlr.de}
\author[4]{\fnm{Elmar} \sur{Brockfeld}}\email{elmar.brockfeld@dlr.de}
\author[3]{\fnm{Oswald} \sur{Knoth}}\email{knoth@tropos.de}
\author[3]{\fnm{Roland} \sur{Schr\"odner}}\email{roland.schroedner@tropos.de}
\author[1,2,5]{\fnm{Jan} \sur{Bumberger}}\email{jan.bumberger@ufz.de}

\affil[1]{\orgdiv{Department of Monitoring and Exploration Technologies}, \orgname{Helmholtz Centre for Environmental Research -- UFZ} \orgaddress{\street{Permoserstra\ss e 15}, \city{Leipzig}, \postcode{04318}, \country{Germany}}}

\affil[2]{\orgdiv{Research Data Management -- RDM}, \orgname{Helmholtz Centre for Environmental Research -- UFZ} \orgaddress{\street{Permoserstra\ss e 15}, \city{Leipzig}, \postcode{04318}, \country{Germany}}}

\affil[3]{\orgdiv{Modeling of Atmospheric Processes}, \orgname{Leibniz Institute for Tropospheric Research (TROPOS)}, \orgaddress{\street{Permoserstra\ss e 15}, \city{Leipzig}, \postcode{04318}, \country{Germany}}}

\affil[4]{\orgdiv{Institute of Transportation Systems}, \orgname{German Aerospace Center (DLR)}, \orgaddress{\street{Rutherfordstra\ss e 2}, \city{Berlin}, \postcode{12489}, \country{Germany}}}

\affil[5]{\orgname{German Centre for Integrative Biodiversity Research (iDiv) Halle-Jena-Leipzig} \orgaddress{\street{Puschstra\ss e 4}, \city{Leipzig}, \postcode{04103}, \country{Germany}}}

\abstract{Road-traffic NO$_2$ hotspots are still often modelled with static emissions and generic temporal profiles, although near-road concentrations respond strongly to rapidly changing traffic conditions. Here, we test whether detector-informed dynamic traffic emissions improve hyperlocal NO$_2$ modelling relative to a conventional static baseline. To this end, we couple an online-calibrated mesoscopic traffic model (SUMO) with the LES-based urban dispersion model CAIRDIO in a nested high-resolution framework for Leipzig, Germany. We compare two otherwise identical experiment setups: a static reference simulation and a coupled simulation in which road-traffic emissions within the SUMO domain are replaced by dynamic emissions derived from simulated traffic states. The framework is designed for city-wide high-resolution application, while the present evaluation focuses on two traffic-oriented hotspot settings during two one-week periods. Compared against hourly NO2 observations of official air quality monitoring, the coupled setup performs better overall, with the clearest improvement at the street-canyon hotspot and in the representation of concentration peaks. Dynamic traffic emissions therefore provide clear added value for hyperlocal NO$_2$ prediction where hotspot realism and exposure-relevant peaks matter.}

\maketitle

\section{Introduction}\label{sec:introduction}

Urban nitrogen dioxide (NO$_2$) pollution remains a persistent concern for public health protection and air-quality management, as ambient concentrations in many European cities do not comply with World Health Organization guidelines  or the more stringent regulatory thresholds set to take effect in 2030 \citep{EEA2024_exceedance_indicator,EEA2025_NO2}. Despite major progress in cleaner combustion technology and the increasing share of electric vehicles in urban fleets, road traffic still contributes a large fraction of urban nitrogen oxide (NO$_x$) emissions \citep{eea2025,pueltz2025}. Tailpipe emissions release NO$_x$ primarily in the form of nitrogen monoxide (NO), which is rapidly converted by ozone in ambient air to the more reactive and highly irritant NO$_2$, the compound that represents the principal health concern of the two. Both short-term and long-term exposure to NO$_2$ have been linked to substantial morbidity, including asthma, chronic lung-function decline, and cardiovascular disease, as well as to premature mortality \citep{Samoli2006,chen2024}. Urban near-field exposure is compounded by the fact that traffic emissions are released directly into the pedestrian environment, increasing the likelihood of immediate inhalation close to the source. Near-road concentration gradients can be steep over the first tens to hundreds of metres from busy roads, which makes a hyperlocal representation particularly relevant for exposure assessment \citep{karner2010}.\\

At the hyperlocal scale, the strongest NO$_2$ gradients emerge where intense traffic activity coincides with inefficient atmospheric ventilation. The vertical mixing of pollutants from pedestrian level is strongly controlled by surrounding building morphology, and pollutant residence times in street canyons can be substantially longer than in open areas \citep{VOORDECKERS20211,Vardoulakis2003}. The formation of urban NO$_2$ hotspots is therefore driven by two interacting controls: microscale flow environments with reduced ventilation, and highly localised traffic dynamics that modulate emissions in space and time \citep{BECKWITH2019,Zhong2016,Dai2022}. Such hotspots frequently occur near signalised intersections, where stop--go conditions, idling, and acceleration phases increase vehicular emissions per unit distance travelled \citep{kim2014}. At somewhat larger scales, the temporal variability of hotspot intensity is additionally shaped by evolving daily traffic demand, network saturation, and congestion episodes \citep{HE2025}. Understanding when and where such hotspots occur is essential for realistic exposure assessment and for the design of targeted mitigation strategies.\\

Because routine air-quality monitoring is spatially sparse and only partly representative of the fine-scale variability in near-road environments, numerical models are commonly used as complementary tools to estimate hyperlocal pollutant distributions \citep{Rodriguez2019,Santiago2022}. Yet the reliability of these models is often limited by uncertainties in meteorological forcing, background composition, urban morphology, and especially emissions \citep{Vardoulakis2004,kim2018,holnicki2015,criado2023}. For road traffic, emissions are still commonly represented through static annual inventories that are spatially disaggregated from areal totals using proxy information such as land use, population density, or road hierarchy \citep{kuenen2014,tno2011,guevara2021}. Traffic counts, if considered at all, often enter only for major roads and rarely capture the complexity of dense urban street networks. As a consequence, static inventory products can exhibit strong local biases in the placement and intensity of traffic emissions, which can propagate into both mesoscale and street-scale air-quality simulations \citep{kuik2018,kim2018}.\\

A further limitation of the conventional inventory approach lies in the temporal representation of emissions. Monthly, weekly, and diurnal variation is usually prescribed through generic temporal factors or gridded temporal-profile products, which may be poorly adapted to regional traffic behaviour and incapable of resolving site-specific patterns in congestion, queue dynamics, or weekday--weekend contrasts \citep{kuenen2014,guevara2021,kumar2021,Poraicu2023,ZHONG2025}. Owing to the uniform application of these time factors and to the fact that real-time traffic information is not included, such approaches inherently struggle to reproduce location-specific temporal characteristics of traffic activity and observed near-road NO$_2$ concentration dynamics \citep{Zhu2018}. This limitation becomes especially critical in high-resolution applications aimed at predicting the magnitude and timing of short-lived concentration peaks, as needed for exposure analysis or for evaluating measures such as low-emission zones and adaptive traffic control. In this context, a useful distinction is between static road-traffic emissions, typically represented by annual inventories combined with standard temporal profiles, and dynamic emissions derived from temporally resolved traffic activity. The latter are, in principle, better suited to capture congestion, signalized stop--go conditions, and other short-lived traffic states that shape exposure-relevant NO$_2$ peaks.\\

Mesoscopic and microscopic traffic simulations offer a plausible route to reduce part of this mismatch between emission representation and hyperlocal exposure modelling. In both approaches, traffic activity is resolved at the level of individual vehicles within a digital road network, with the main difference lying in vehicle trajectory computation. While  microscopic simulations accurately resolve vehicle interactions through very small integration time steps \citep{gips1981,treiber2000}, mesoscopic models follow an event-based approach (e.g., when a vehicle reaches a junction) for updating vehicle movements \citep{Horni2016}. This better integration efficiency makes mesoscopic models more adapted for representing large-scale networks of, e.g., entire cities. Still, in both approaches local traffic activity is informed by temporally resolved traffic demands, and where available, traffic observations \citep{Lopez2018,ALGHERBAL2025}, making them equally useful not only in transport-system analysis and urban planning \citep{Soni2023,Schaffland2024}, but also traffic emission estimation \citep{Madziel2023,Jang2025}.
By combining simulated traffic states with vehicle- and situation-specific emission factor systems such as Handbook Emission Factors for Road Transport (HBEFA) \citep{hbefa42}, one can derive spatially and temporally resolved network-wide emissions that respond to changing traffic conditions \citep{Madziel2023,kim2020,Jang2025}. For real-time or near-real-time applications, however, the reliability of such estimates depends critically on the availability and quality of traffic flow detector data for online calibration as well as on the calibration strategy itself \citep{kim2020,Keler2023}.\\

At the same time, near-road NO$_2$ hotspots are not governed by emissions alone, but by the interaction of traffic dynamics with street-canyon aerodynamics, urban geometry, non-local background contributions, turbulent exchange, and rapid NO--NO$_2$--O$_3$ chemistry \citep{Zhong2016,Lateb2016,weger2021,weger2022,Dai2022}. Accordingly, only a limited number of studies have embedded traffic-simulation-informed emissions into urban digital-twin or end-to-end urban air-quality frameworks for operationally relevant applications, despite recent advances in high-resolution urban air-quality modelling systems \citep{kim2018,Maronga2020,Khan2021,Resler2021,Samad2024}. Existing workflows are often based on semi-idealised or simplified setups that do not combine detector-informed traffic simulation with realistic evolving meteorological and chemical boundary conditions \citep{ESSAMLALI2025,Laudan2025}. Notable exceptions include the digital-twin study of \citet{Ilarri2022} for Zaragoza and the nested mesoscale--microscale case study of \citet{SANJOSE2021} for Madrid. Yet, while these studies demonstrate technical feasibility, they do not directly quantify the added value of dynamic traffic simulations for urban air-quality prediction against a conventional static-emission reference. As a result, it remains insufficiently understood whether dynamic, traffic-informed emission representations measurably improve hyperlocal NO$_2$ modelling compared with a conventional static baseline. Beyond the immediate attribution question, this distinction is also relevant for urban digital-twin applications in which hyperlocal air-quality model fidelity conditions the credibility of downstream scenario analysis, predictive assessment, and environmentally sensitive traffic-management strategies and urban planning processes.\\

This study addresses that gap through a hypothesis-driven comparison between two otherwise identical modelling setups for Leipzig, Germany. In the control experiment (CTR), road traffic is represented by static inventory-based emissions combined with standard temporal factors. In the coupled experiment (CPL), the static road-traffic sector is replaced within the SUMO model domain by dynamic emissions derived from simulated traffic states. The study builds on a high-resolution urban modelling framework that couples the online-calibrated traffic model SUMO \citep{Lopez2018} with the LES-based dispersion model CAIRDIO \citep{weger2021}. Particular emphasis is placed on physical realism: a multi-domain nesting configuration driven by evolving external meteorological and air-composition boundary conditions from mesoscale models is used to represent turbulent transport and chemistry at street level \citep{weger2022}. While the present analysis focuses on selected hotspot-relevant locations and episodes, the coupled modelling framework is not limited to such local application windows, but is designed to resolve urban NO$_2$ fields at comparable spatial detail across the entire city area. We evaluate the simulations against official NO$_2$ monitoring observations over two one-week periods in early April (T1) and early November 2025 (T2), with particular emphasis on hotspot behaviour and peak performance. The aim is therefore not merely to present a coupled workflow, but to quantify the added value and limitations of dynamic traffic emissions for reproducing observed urban NO$_2$ hotspots. In this sense, the study addresses a prerequisite to support decision-making based on urban digitial twins, namely whether the underlying hyperlocal NO$_2$ model responds meaningfully to dynamic traffic information at the scales relevant for hotspot management.\\

The study is guided by three research questions:
\begin{enumerate}
    \item Do dynamic road-traffic emissions improve agreement with observed hourly NO$_2$ time series compared with a static traffic-emission baseline?
    \item Is the added value larger at a street-canyon hotspot than at a more open traffic site?
    \item Does the dynamic approach improve exposure-relevant peak characteristics, in particular peak amplitude and peak timing?
\end{enumerate}


\section{Results}\label{sec:results}

\subsection{Dynamic traffic emissions alter the spatial and temporal pattern of road-traffic NO$_x$ emissions}

\begin{figure}[H]
\vspace*{2mm}
\begin{center}
\includegraphics[width=\textwidth]{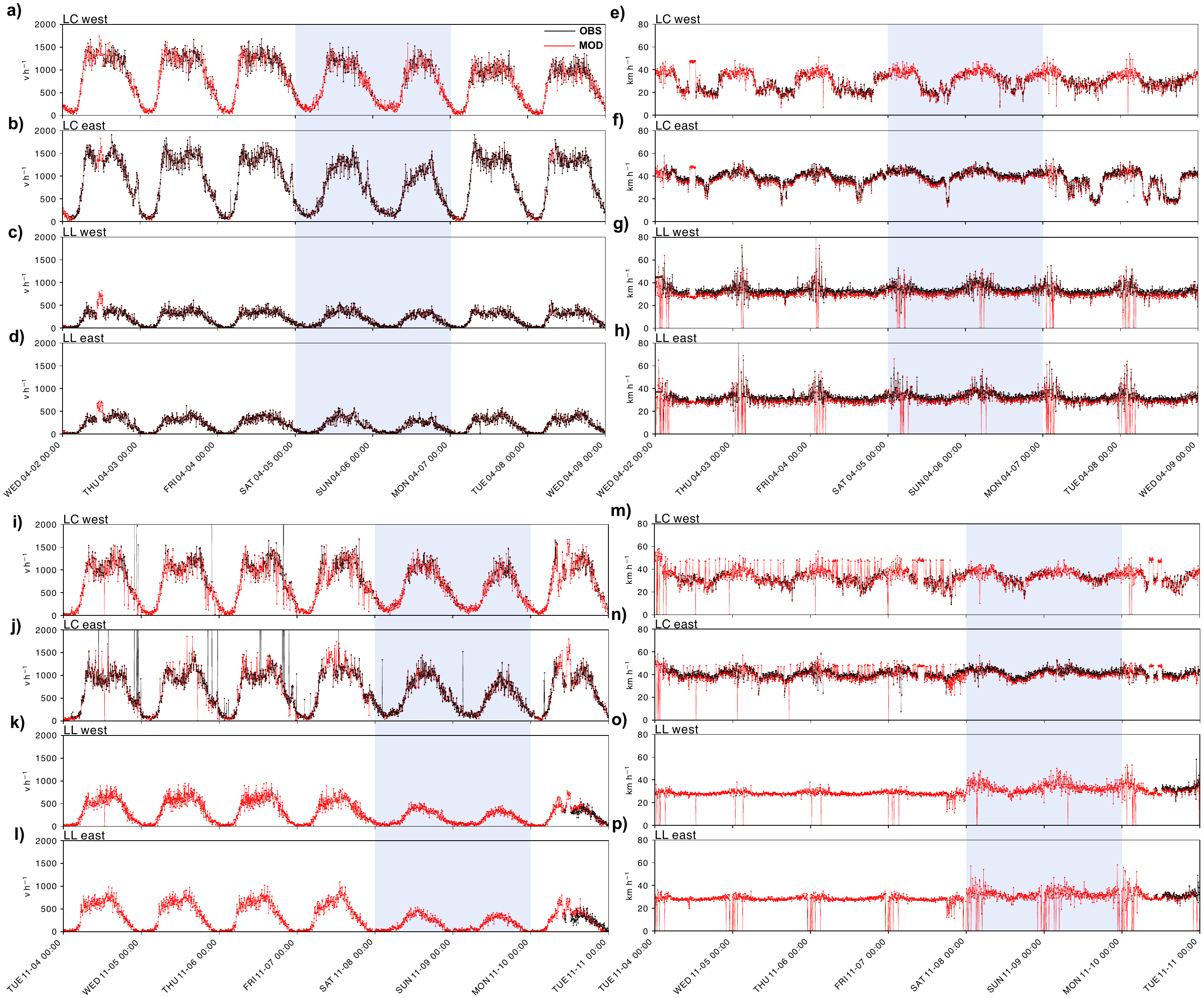}
\end{center}
\caption{Panels (a--d): Time series of modelled (red) and observed (black) traffic flows at the two calibrator pairs (westward and eastward lane directions) in the vicinity of Leipzig Center (LC; CAIRDIO domain L3) and Leipzig L\"utzner Stra\ss e (LL; CAIRDIO domain L4) during study period T1. Panels (e--h): Corresponding modelled and observed average traffic speeds. Panels (i--l): Same as panels (a--d), but for study period T2. Panels (m--p): Same as panels (e--h), but for study period T2. Saturdays and Sundays are further highlighted by light-blue shading.}
\label{fig:sumo_vs_det_time_series}
\end{figure}

\begin{figure}[H]
\vspace*{2mm}
\begin{center}
\includegraphics[width=\textwidth]{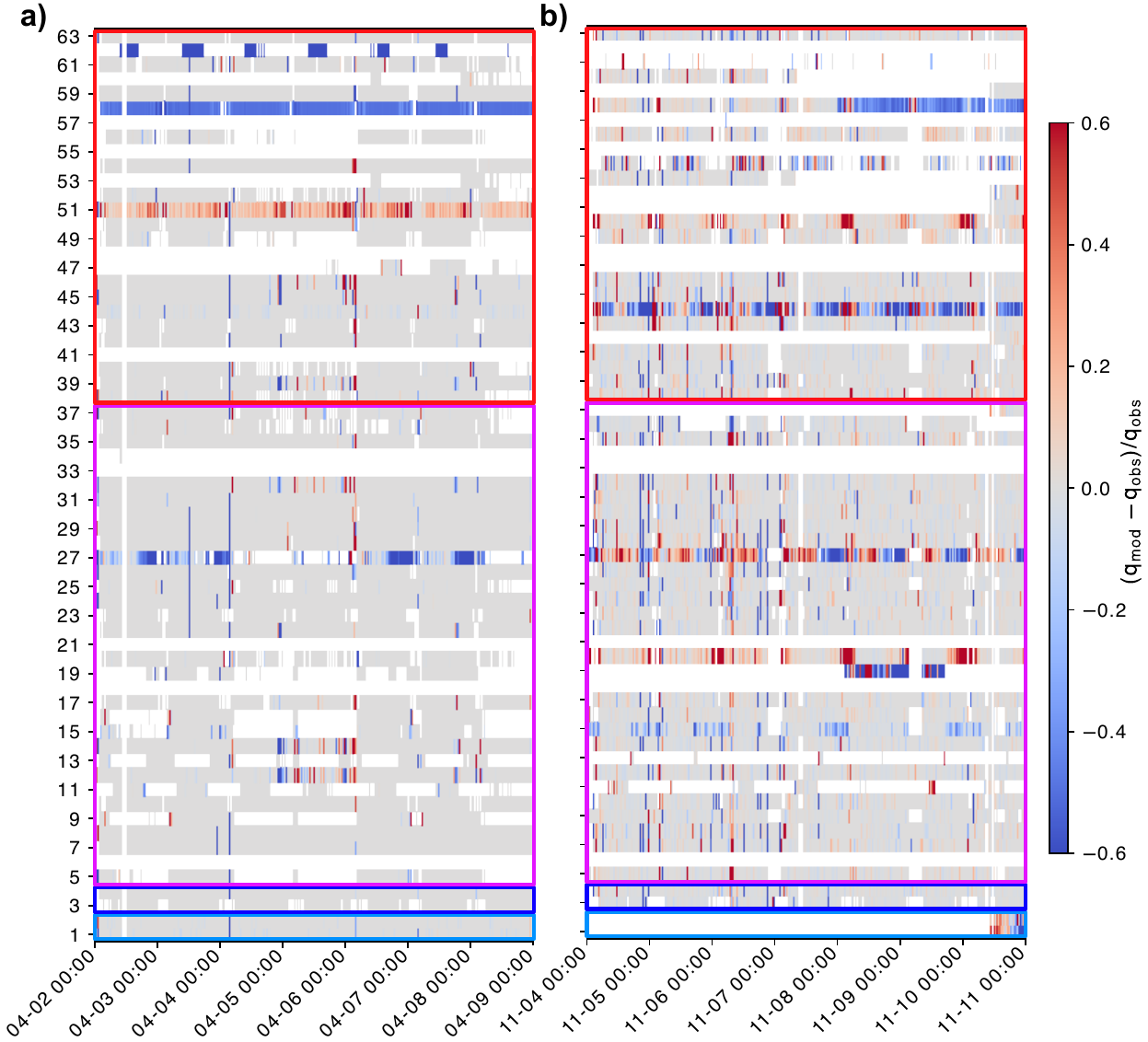}
\end{center}
\caption{Heatmaps of the relative deviation between modelled and observed traffic volumes (vehicles per hour, $\unit{v\,h^{-1}}$) at all 63 calibration sites during validation period T1 (a) and T2 (b). Displayed data intervals are \SI{5}{min}. Missing values are indicated in white. Detector sites are grouped according to their primary association with domains L4 (light blue frame), L3 (dark blue frame), L2 (violet frame), and L1 (red frame).}
\label{fig:sumo_calibration_q}
\end{figure}

\begin{figure}[H]
\vspace*{2mm}
\begin{center}
\includegraphics[width=\textwidth]{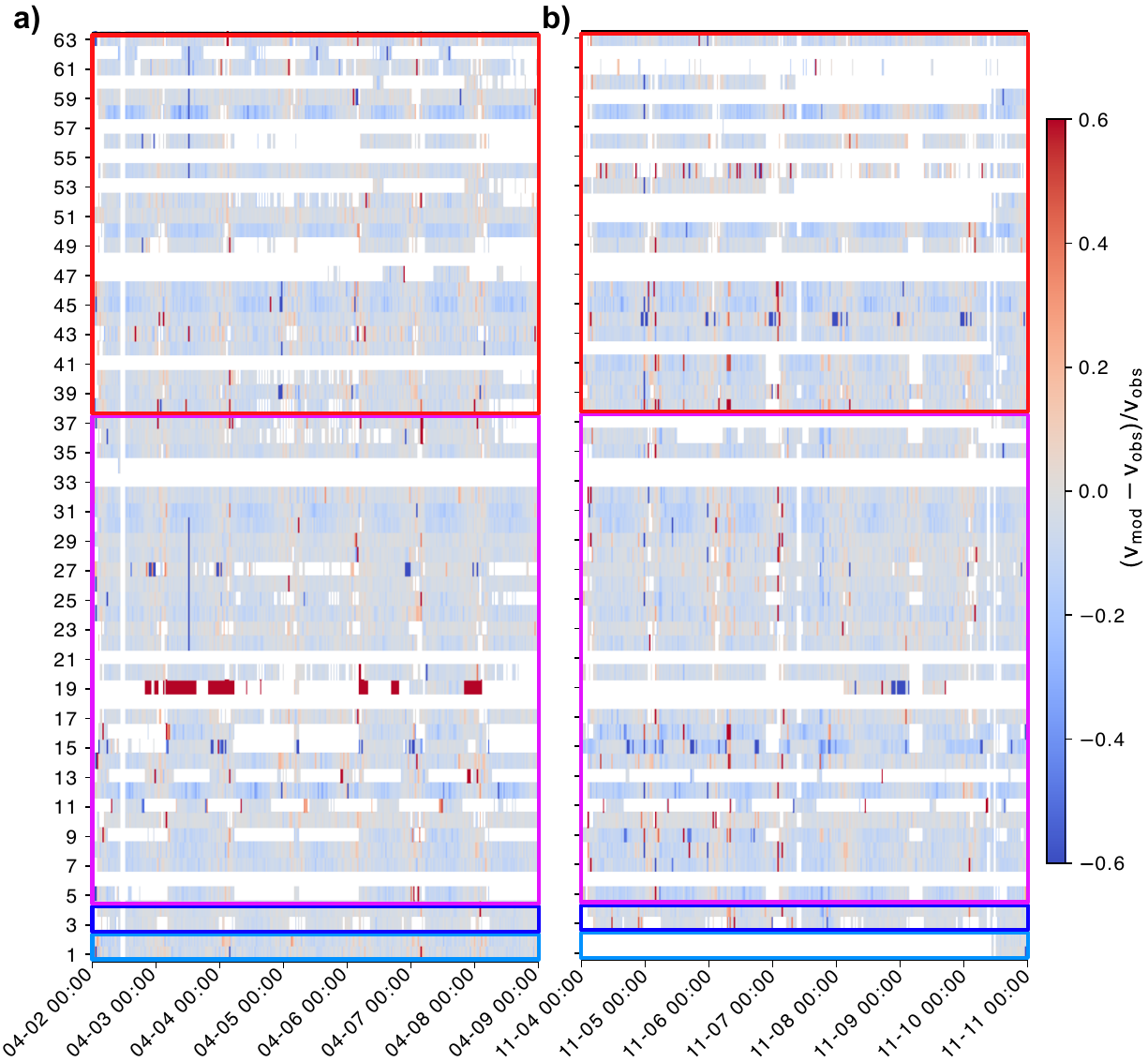}
\end{center}
\caption{Heatmaps of the relative deviation between modelled and observed traffic speeds ($\unit{km\,h^{-1}}$) at all 63 calibration sites during validation period T1 (a) and T2 (b). Displayed data intervals are \SI{5}{min}. Missing values are indicated in white. Detector sites are grouped according to their primary association with domains L4 (light blue frame), L3 (dark blue frame), L2 (violet frame), and L1 (red frame).}
\label{fig:sumo_calibration_v}
\end{figure}

The dynamic road-traffic emissions in CPL are based on detector-informed SUMO simulations that reproduce the observed traffic state with generally good accuracy. As shown by time series of traffic volumes $q$ (vehicles per hour, \SI{}{v\,h^{-1}}) and speeds $v$ (\SI{}{\kilo\meter\per\hour}) in Figure~\ref{fig:sumo_vs_det_time_series}, there is generally a very good overlap between modelled and observed traffic characteristics whenever detector data are available. This is quantitatively confirmed by the calibration heatmaps in Figures \ref{fig:sumo_calibration_q} and \ref{fig:sumo_calibration_v}. Overall, relative deviations remain below $5\,\unit{\%}$ at most calibration sites employed in SUMO. Larger deviations occur only sporadically and are mainly associated with rapidly fluctuating counts or with erroneous detector reports. Across all calibration sites, the fractional bias (FB) in traffic volume is small in both evaluation periods, ranging from $-0.03$ in T1 to $-0.06$ in T2, while the root-mean-square error amounts to $16\,\unit{v\,h^{-1}}$ and $25\,\unit{v\,h^{-1}}$ for T1 and T2, respectively. Simulated traffic speeds show a comparable overall level of agreement, with relative deviations below $10\,\unit{\%}$ for most reported values, although a slight systematic negative daytime bias is apparent at many sites ($\mathrm{FB}=-0.05$ for T1 and $\mathrm{FB}=-0.07$ for T2 across all sites). Overall, these results indicate that the dynamic emission fields used in CPL are grounded in a realistically calibrated representation of urban traffic dynamics.

\begin{figure}[H]
\vspace*{2mm}
\begin{center}
\includegraphics[width=1.0\textwidth]{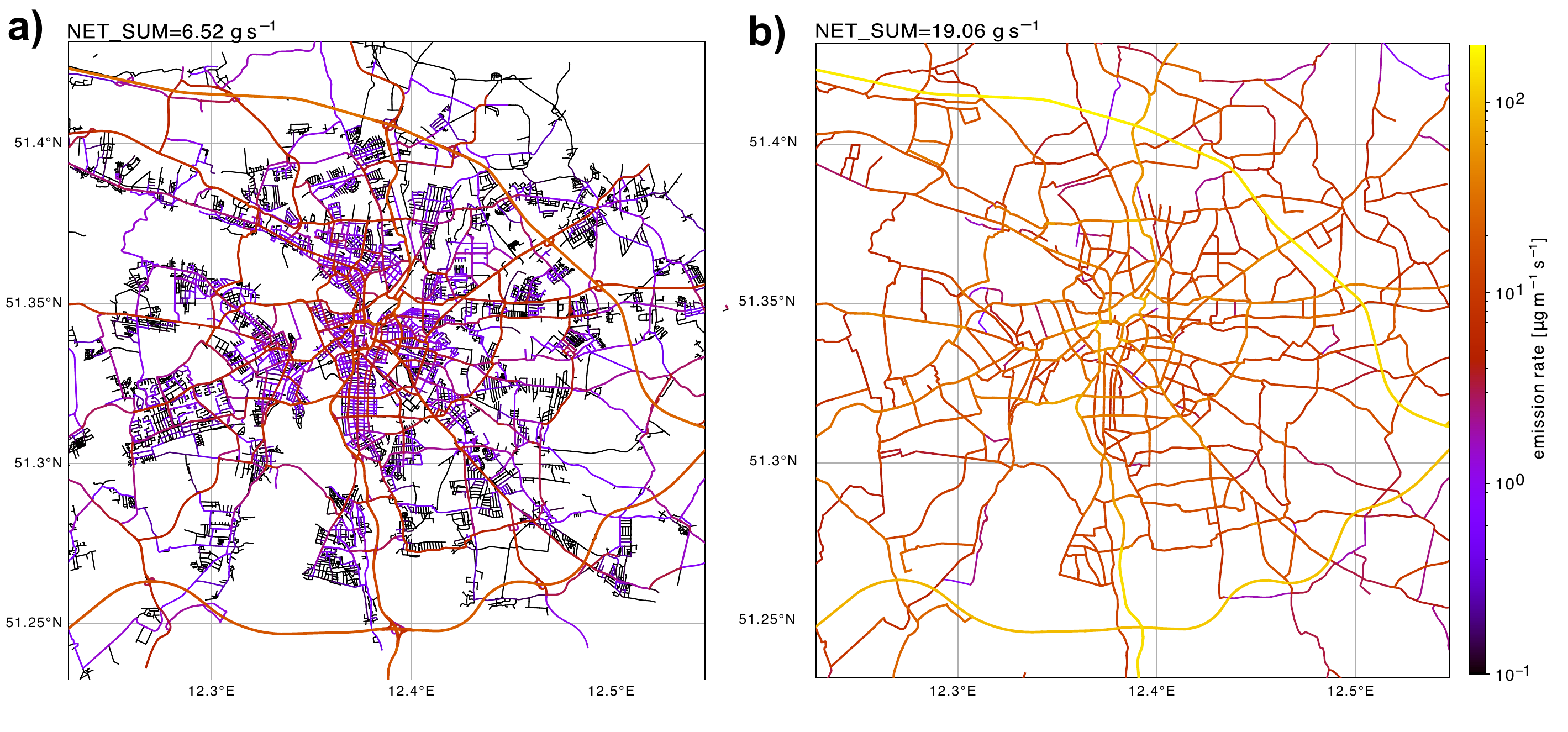}
\end{center}
\caption{Road transport (SNAP 7) NOx emissions within domain L1 averaged over the period 4 - 12 Nov 2025 (from 00:00 UTC to 00:00 UTC): a) shows the dynamic SUMO emissions and b) the static emissions after application of a reduction factor of 0.3 motivated in Section \ref{sec:emissions}. Note that the SUMO model does not cover the entire area of domain L1, so the lower right and upper right corners in (b) are left blank. The numbers at the top left of the panels show respective cumulative emissions within domain L1. }
\label{fig:traffic_emissions_L1}
\end{figure}

\begin{figure}[H]
\vspace*{2mm}
\begin{center}
\includegraphics[width=0.6\textwidth]{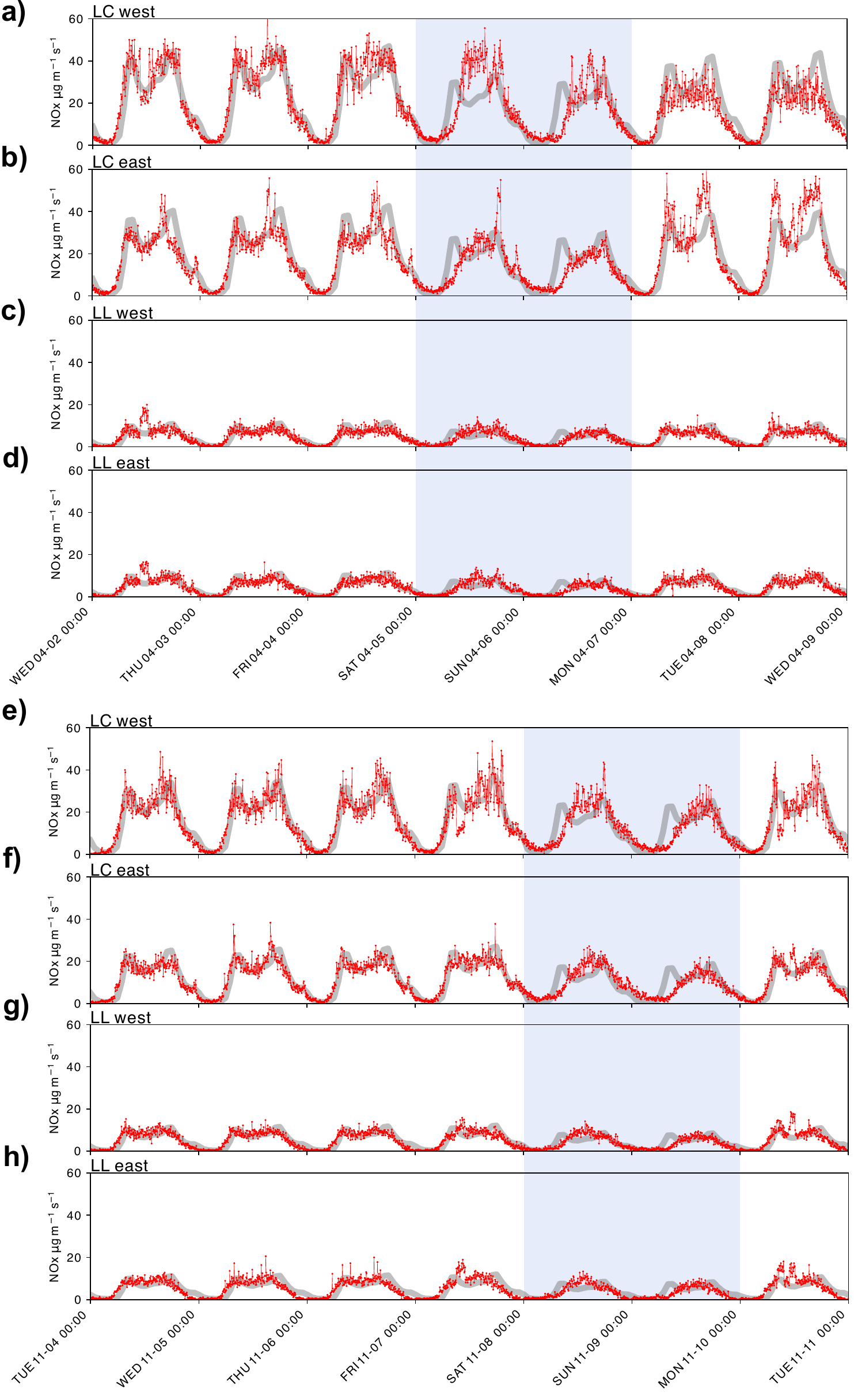}
\end{center}
\caption{Panels (a-d): Time series of edge-based SUMO NO$_x$ emission rates (red graphs) at the two calibrator pairs (westward \& eastward lane directions) in  the vicinity of Leipzig Center (LC) site (CAIRDIO domain L3), and Leipzig Lützner Str. (LL) site (CAIRDIO domain L4) during study period T1. For comparison, the solid grey line in the background shows the SNAP7 emission profile used in the static emission approach \citep{tno2011} that is further scaled to match the temporal mean of respective SUMO emission rate. Panels (e-h): Same as panels (a-d), but for study period T2. Saturdays and Sundays are further highlighted by light-blue shading.}
\label{fig:edge_emissions_time_series}
\end{figure}

The calibrated traffic dynamics translate into marked differences between the dynamic and static road-traffic emission representations, both in their spatial distribution across the urban road network (Figure~\ref{fig:traffic_emissions_L1}) and in their local temporal behaviour (Figure~\ref{fig:edge_emissions_time_series}). Figure~\ref{fig:traffic_emissions_L1} shows that the spatial distribution of NO$_x$ emissions in CPL differs substantially from the static inventory-based representation, both in local intensity and in the allocation of emissions within the urban road network, which is represented in greater detail in SUMO. These differences indicate that the dynamic setup captures locally varying traffic activity that is not resolved by the static baseline.\\

At the local scale, road-edge-based emission time series (units of \SI{}{\micro\gram\per\meter\per\second}) further highlight the contrasting temporal behaviour of the two approaches (Figure~\ref{fig:edge_emissions_time_series}). While the static setup follows the prescribed SNAP7 temporal profile by construction, the SUMO-derived emissions respond strongly to locally evolving traffic states. The largest deviations occur during periods influenced by congestion and generally during weekends when the diurnal pattern in traffic demands substantially differs from weekday conditions. This shows that the dynamic representation is better suited to capture short-lived traffic states that are relevant for near-road NO$_2$ hotspots and exposure peaks.

\subsection{The coupled setup reproduces urban NO$_2$ dynamics at city and street scales}

\begin{figure}[H]
\vspace*{2mm}
\begin{center}
\includegraphics[width=1.0\textwidth]{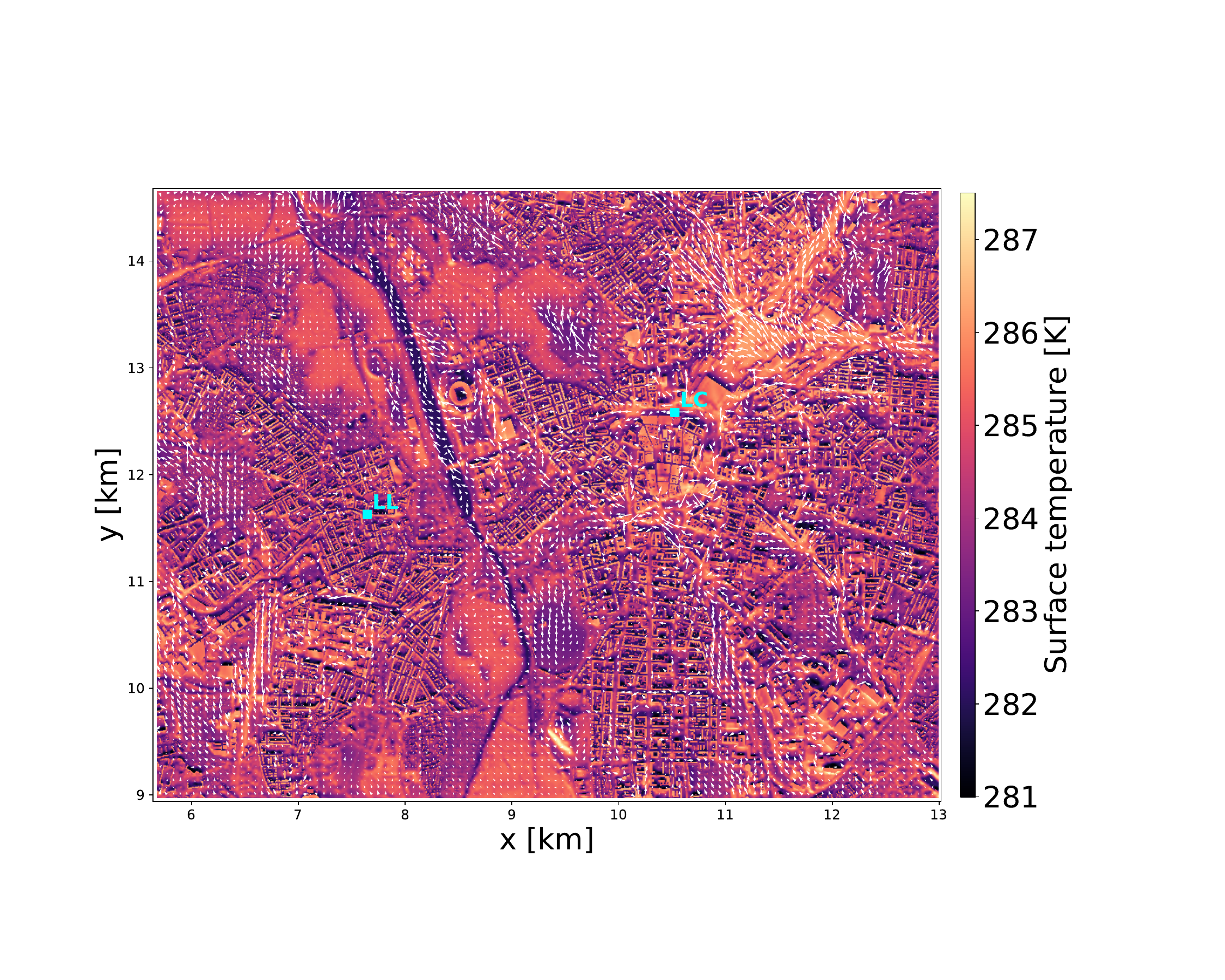}
\end{center}
\caption{Depiction of the simulated urban wind field (white quivers) driving the dispersion simulation within domain L2 ($20\,\unit{m}$ resolution) on 5 November at 16:00 local time. The colour shading shows simulated surface temperature, illustrating the combined influence of urban morphology, land cover, and thermal contrasts on the local flow field. The target sites LC and LL are further depicted by cyan markers and labels.}
\label{fig:cairdio_l2_tsurf}
\end{figure}

A realistic representation of the urban meteorological environment is a prerequisite for credible street-scale NO$_2$ simulations. In CAIRDIO, the LES-based flow simulation resolves spatially heterogeneous wind and temperature patterns that respond to urban morphology, surface properties, and transient boundary forcing. Figure~\ref{fig:cairdio_l2_tsurf} illustrates physically plausible microscale structure in the near-surface wind and thermal fields across domain L2, including reduced wind speeds in densely built-up areas and locally enhanced flow along more exposed or channelled surfaces.\\

\begin{figure}[H]
\vspace*{2mm}
\begin{center}
\includegraphics[width=1.0\textwidth]{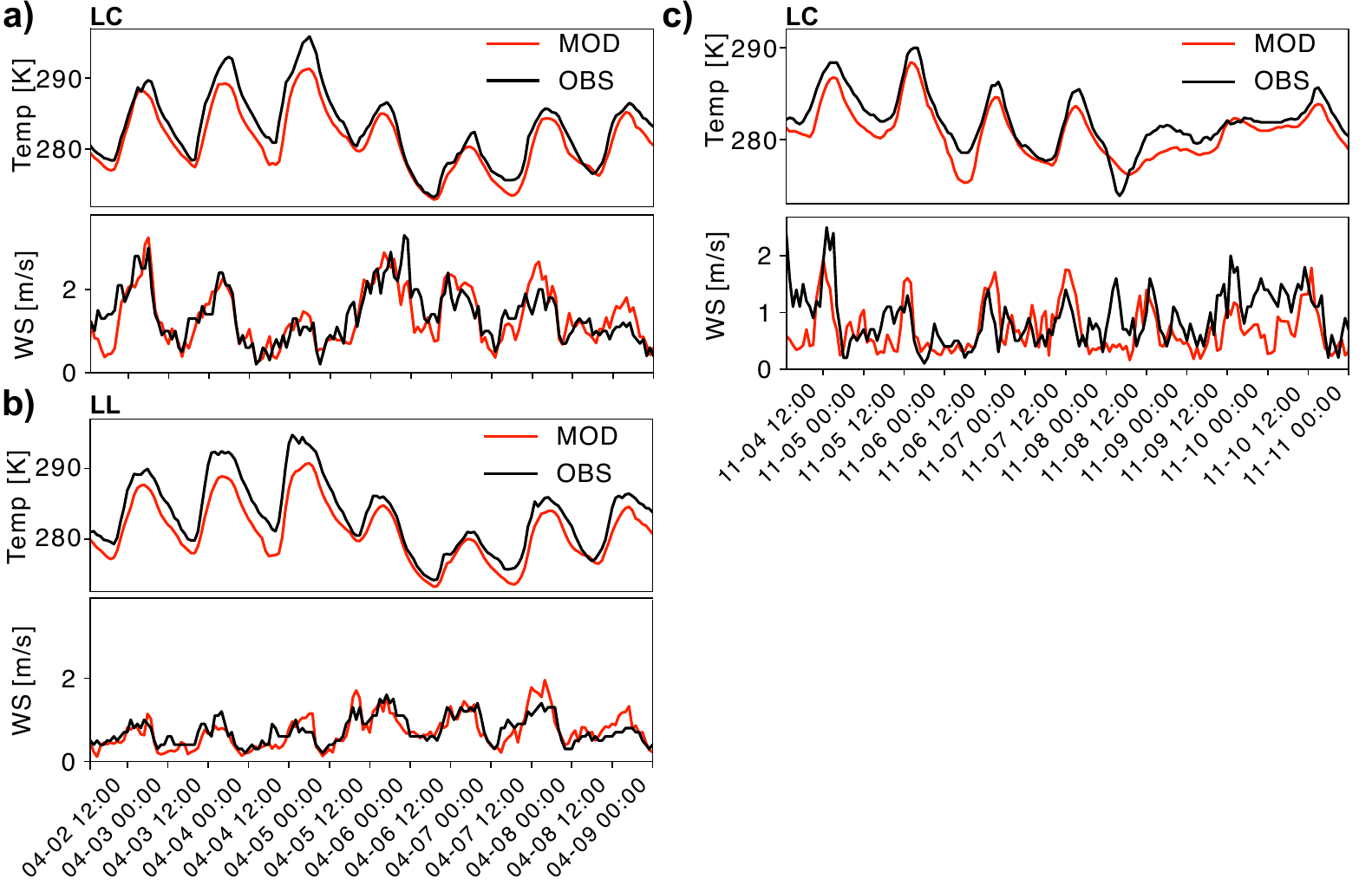}
\end{center}
\caption{Time series of modelled and observed air temperature and wind speed at site LC during period T1 (a), LL during period T1 (b), and LC during period T2 (c).}
\label{fig:temp_ws_graphs_lpzmitte}
\end{figure}

This qualitative realism is supported by comparison with site observations. As shown in Figure~\ref{fig:temp_ws_graphs_lpzmitte}, the model generally captures the observed temporal evolution of air temperature and wind speed during the evaluation periods, including the main synoptic and diurnal trends. In particular, the diurnal temperature cycle during the first days of T1 is reproduced well, while the subsequent damping under persistent cloud cover is also represented, although the onset of cloud effects appears slightly too early in the model. Overall, these findings indicate that the meteorological component of the coupled setup provides a credible basis for the simulated urban NO$_2$ fields.

\begin{figure}[H]
\vspace*{2mm}
\begin{center}
\includegraphics[width=1.0\textwidth]{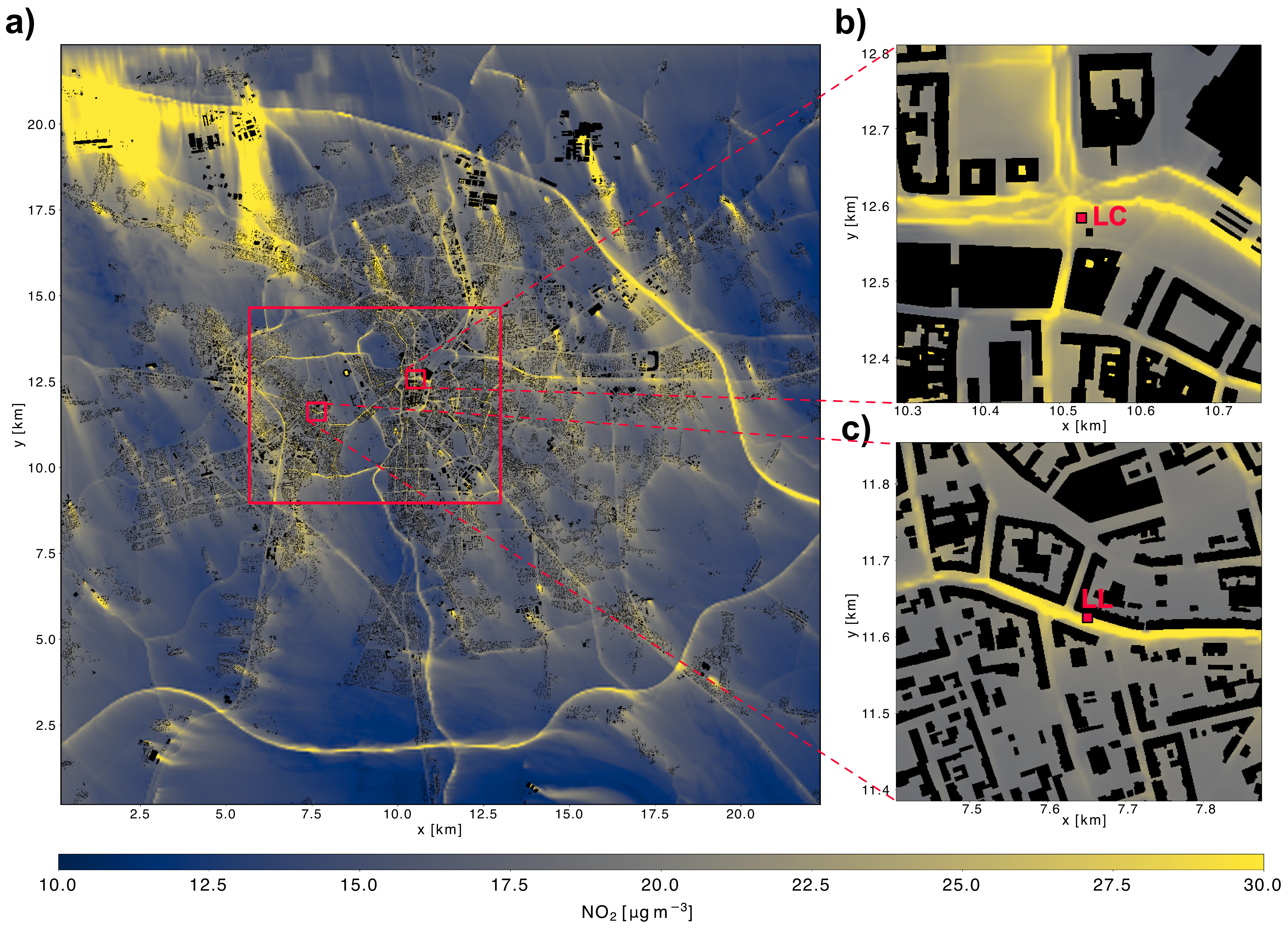}
\end{center}
\caption{Simulated hourly-mean near-surface NO$_2$ concentration of the coupled model run CPL within (a) domains L1 and L2 (marked by outer red frame) and (b) and (c) within domains L3 and L4, respectively at 4 April 08:00 – 09:00 local time. Building footprints are shown in black colour.}
\label{fig:cairdio_overview_no2_conc}
\end{figure}

The resulting NO$_2$ fields show that the coupled setup resolves concentration variability from the city-scale down to street-scale hotspots. Figure~\ref{fig:cairdio_overview_no2_conc}a illustrates the simulated near-surface NO$_2$ distribution during a morning rush-hour period, highlighting major traffic corridors and elevated concentrations downwind of the main urban source regions. Besides traffic emissions, the city-scale pattern also reflects contributions from other sectors, namely industrial and residential sources described through the static emission inventory, and their modulation by the prevailing wind field. At this scale, major roads and highways coincide with pronounced NO$_2$ hotspots, while elevated urban background concentrations emerge from the dispersion of both spatially focused and diffuse area sources.\\

At finer resolution, Figures~\ref{fig:cairdio_overview_no2_conc}b and \ref{fig:cairdio_overview_no2_conc}c demonstrate that the model captures pronounced street-scale hotspot structures near the two monitoring sites. The simulated hotspot at Lützner Straße (LL) is more strongly amplified and spatially confined, consistent with its street-canyon setting and reduced ventilation, whereas Leipzig Center (LC) is embedded in a comparatively more open environment. These results provide qualitative evidence that the coupled SUMO--CAIRDIO framework reproduces both city-wide NO$_2$ patterns and the local concentration structures relevant for hotspot analysis.

\subsection{Dynamic emissions improve overall agreement with observed NO$_2$}

\begin{figure}[H]
\vspace*{2mm}
\begin{center}
\includegraphics[width=1.0\textwidth]{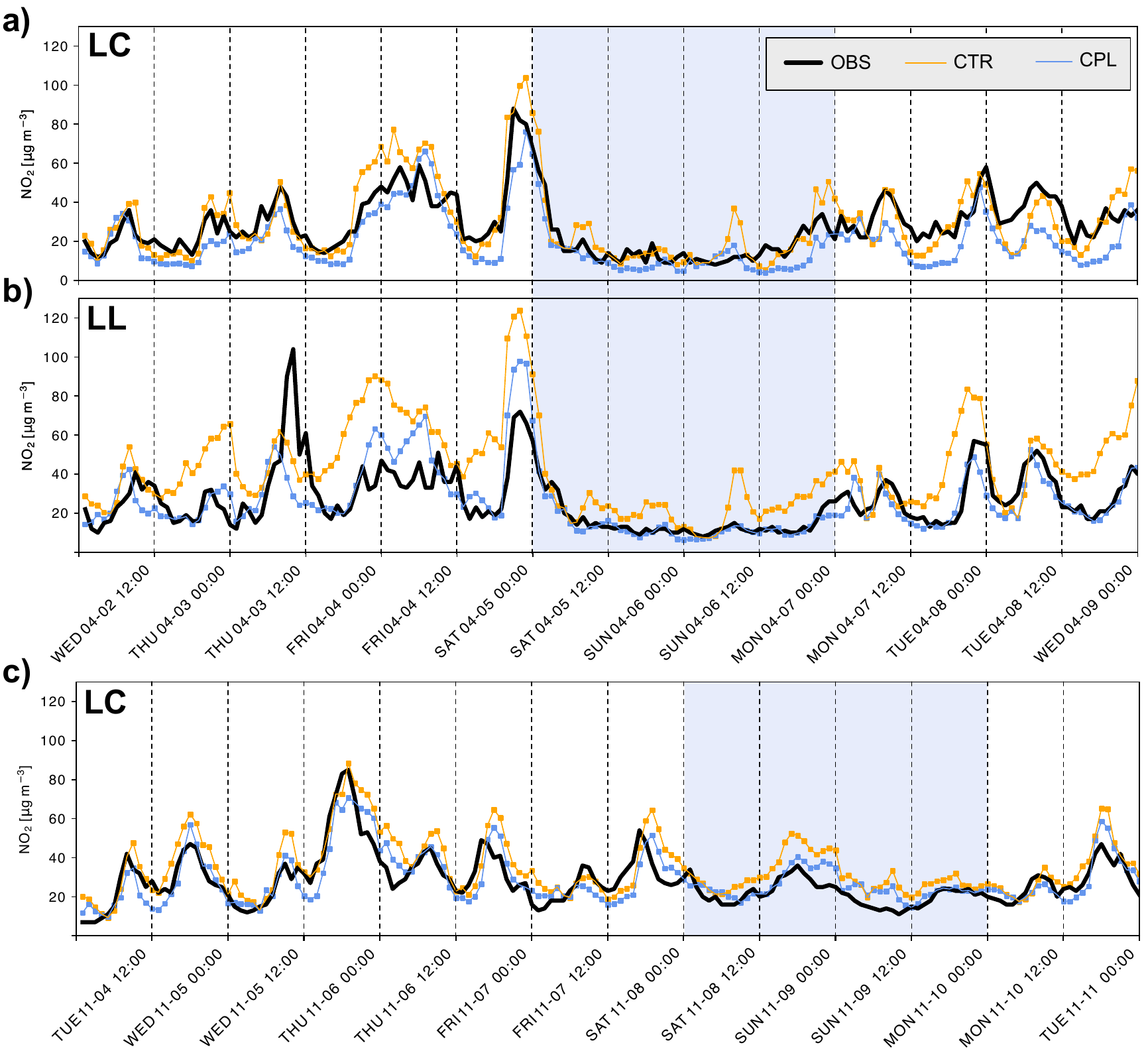}
\end{center}
\caption{Observed and modelled hourly NO$_2$ time series at (a) Leipzig Center (LC) and (b) Leipzig Lützner Straße (LL) during validation period T1 in April 2025, and at (c) LC during validation period T2 in November 2025. Observations are shown in black, the static reference simulation CTR in orange, and the coupled simulation CPL in blue. Saturdays and Sundays are further highlighted by light-blue shading.}
\label{fig:no2_time_series}
\end{figure}

Hourly NO$_2$ time series at the monitoring sites LC and LL are shown in Figure~\ref{fig:no2_time_series}. For the November period T2, only site LC is available, so that three one-week evaluation series are considered in total. Across both sites and periods, the observations show a consistent temporal structure, with pronounced weekday peaks reaching up to $100\,\unit{\mu g\,m^{-3}}$, lower concentrations during weekends, and clear morning and evening rush-hour enhancements, especially during T2. The spring period T1 differs in that stronger photochemical depletion during daytime shifts the daily minima towards the early afternoon, while the highest NO$_2$ concentrations tend to occur during nighttime due to replenishment from the reaction of NO with O$_3$.\\

Against these observations, the coupled experiment CPL generally reproduces the main temporal features well (Figure~\ref{fig:no2_time_series}). In particular, the weekday peak structure and the flatter weekend behaviour are represented realistically across the evaluated series. A slight systematic underestimation is visible at site LC during T1, but the overall temporal evolution remains well captured. The static reference experiment CTR shows a less consistent behaviour. During T1, the static traffic emissions are clearly higher than the corresponding SUMO-derived emissions, which leads to an apparent improvement at LC but to a substantial overestimation at LL. During T2, the differences between CPL and CTR are smaller overall, yet CTR still tends to overestimate NO$_2$, especially during the weekend.\\

\begin{figure}[H]
\vspace*{2mm}
\begin{center}
\includegraphics[width=0.6\textwidth]{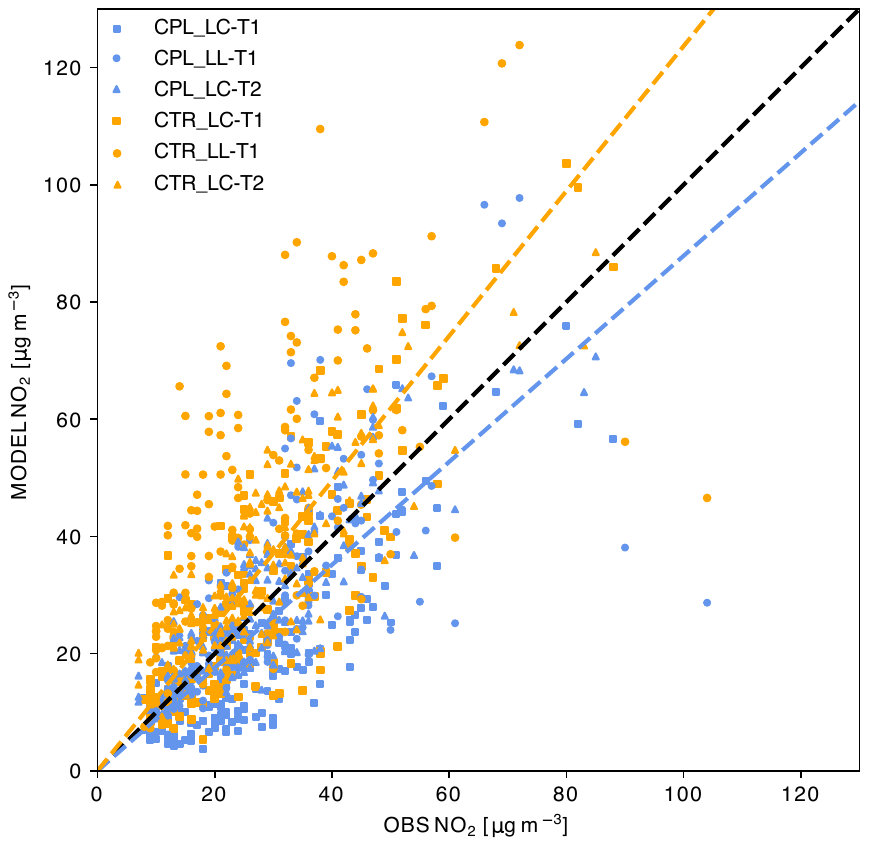}
\end{center}
\caption{Scatter plot of the observed vs. modelled hourly NO$_2$ time series shown in Figure~\ref{fig:no2_time_series}. LC-T1 is shown by squares, LL-T1 by circles, and LC-T2 by triangles. Blue markers show CPL and orange markers CTR data points. Additionally shown are the lines of equality (black dashed line), and of the zero-intercept regressions to the CPL (blue dashed line) and CTR results (orange dashed line).}
\label{fig:no2_scatter_plot}
\end{figure}

The scatter plots of the combined model series presented in Figure~\ref{fig:no2_scatter_plot} provide a more conclusive picture in this regard, confirming the general overestimation of NO$_2$ in CTR, which is particularly pronounced during peak NO$_2$ episodes. In contrast, CPL exhibits a more balanced performance, with a less pronounced overall underestimation of NO$_2$. This generally better agreement of CPL is quantitatively confirmed by the summary metrics in Table~\ref{tab2}. Averaged across all evaluated time series, CPL outperforms CTR in all performance measures, with a lower fractional bias ($-0.10$ vs.\ $0.26$), lower centered root-mean-square error ($10.49$ vs.\ $14.17\,\unit{\mu g\,m^{-3}}$), higher coefficient of determination ($0.45$ vs.\ $-0.28$), and higher index of agreement ($0.86$ vs.\ $0.77$). The strongest difference occurs at LL during T1, where CTR fails to reproduce the observations and even yields a negative $\mathrm{R}^2$. Note that negative $\mathrm{R}^2$ values are possible, particularly in applications beyond standard linear regression settings, and indicate a model performance worse than that of a mean predictor. In contrast to CTR, CPL maintains near-zero bias and substantially better overall agreement at LL. At LC during T1, by contrast, CTR performs slightly better in several bulk metrics, consistent with the moderate underestimation visible in CPL.\\

Taken together, these results show that SUMO-derived dynamic traffic emissions lead to a more robust and spatially consistent NO$_2$ simulation than the static-emission baseline, even though the magnitude of the improvement depends on site and period.

\begin{table}[h]
\caption{General agreement metrics for the NO$_2$ observation and model time series shown in Figure~\ref{fig:no2_time_series}. FB denotes fractional bias, CRMSE the centered root-mean-square error, $\mathrm{R}^2$ the coefficient of determination, and IOA Willmott's index of agreement.}
\label{tab2}
\begin{tabular*}{\textwidth}{@{\extracolsep\fill}lcccccccc}
\toprule
& \multicolumn{4}{@{}c@{}}{CPL} & \multicolumn{4}{@{}c@{}}{CTR} \\\cmidrule{2-5}\cmidrule{6-9}
Time series & FB & CRMSE\footnotemark[1] & $\mathrm{R}^2$ & IOA & FB & CRMSE\footnotemark[1] & $\mathrm{R}^2$ & IOA \\
\midrule
LC-T1  & -0.36 & \textbf{8.35} & 0.35 & 0.84 & \textbf{0.06} & 10.58 & \textbf{0.48} & \textbf{0.90}\\
LL-T1  & \textbf{-0.01} & \textbf{12.25} & \textbf{0.39} & \textbf{0.84} & 0.47 & 17.66 & -1.34 & 0.65\\
LC-T2  & \textbf{0.03} & \textbf{7.59} & \textbf{0.65} & \textbf{0.91} & 0.23 & 8.20 & 0.29 & 0.85\\
\midrule
Total  & \textbf{-0.10} & \textbf{10.49} & \textbf{0.45} & \textbf{0.86} & 0.26 & 14.17 & -0.28 & 0.77\\
\botrule
\end{tabular*}
\footnotetext[1]{in units of \SI{}{\micro\gram\per\cubic\meter}}
\end{table}

\subsection{Added value is largest at the street-canyon hotspots}

The benefit of dynamic traffic emissions is not spatially uniform across the evaluated sites. The clearest improvement occurs at Leipzig Lützner Straße (LL), where the static reference simulation CTR shows a pronounced positive bias during T1, while the coupled simulation CPL reproduces both the magnitude and temporal evolution of observed NO$_2$ substantially better (Figure~\ref{fig:no2_time_series}b; Table~\ref{tab2}). This contrast is particularly evident during weekday peak periods, for which CTR tends to overestimate concentrations persistently, whereas CPL remains much closer to the observations.\\

While the added value of CPL at LL may be coincidental, it is still consistent with the local setting of the site. As indicated by the concentration fields in Figure~\ref{fig:cairdio_overview_no2_conc}, LL is embedded in a street-canyon environment (width $\sim$ \SI{17}{m}) with stronger confinement and hotspot amplification, while Leipzig Center (LC) is situated in a comparatively more open setting. Under the canyon conditions at LL, the interaction between locally varying traffic states and reduced ventilation appears to make the NO$_2$ simulation more sensitive to how road-traffic emissions are represented. In this respect, the dynamic SUMO-derived emissions provide a more realistic description of the short-term traffic conditions that shape near-road accumulation.\\

At LC, by contrast, the difference between CPL and CTR is smaller. During T1, CTR even performs slightly better in several bulk agreement metrics (Table~\ref{tab2}), despite its less realistic traffic-emission representation. This suggests that in the more open environment at LC, the influence of nearby traffic emissions is less dominant and more likely offset by other uncertainties, including remote urban emissions and background concentrations. During T2, CPL again shows the more balanced behaviour, particularly by reducing the weekend overestimation visible in CTR (Figure~\ref{fig:no2_time_series}c).\\

Taken together, these results indicate that the added value of dynamic traffic emissions is largest at the street-canyon hotspot, where locally evolving traffic conditions interact most strongly with reduced ventilation and near-road accumulation.\\

\subsection{Dynamic emissions improve peak magnitude and reduce false hotspot peaks}

\begin{figure}[H]
\vspace*{2mm}
\begin{center}
\includegraphics[width=1.0\textwidth]{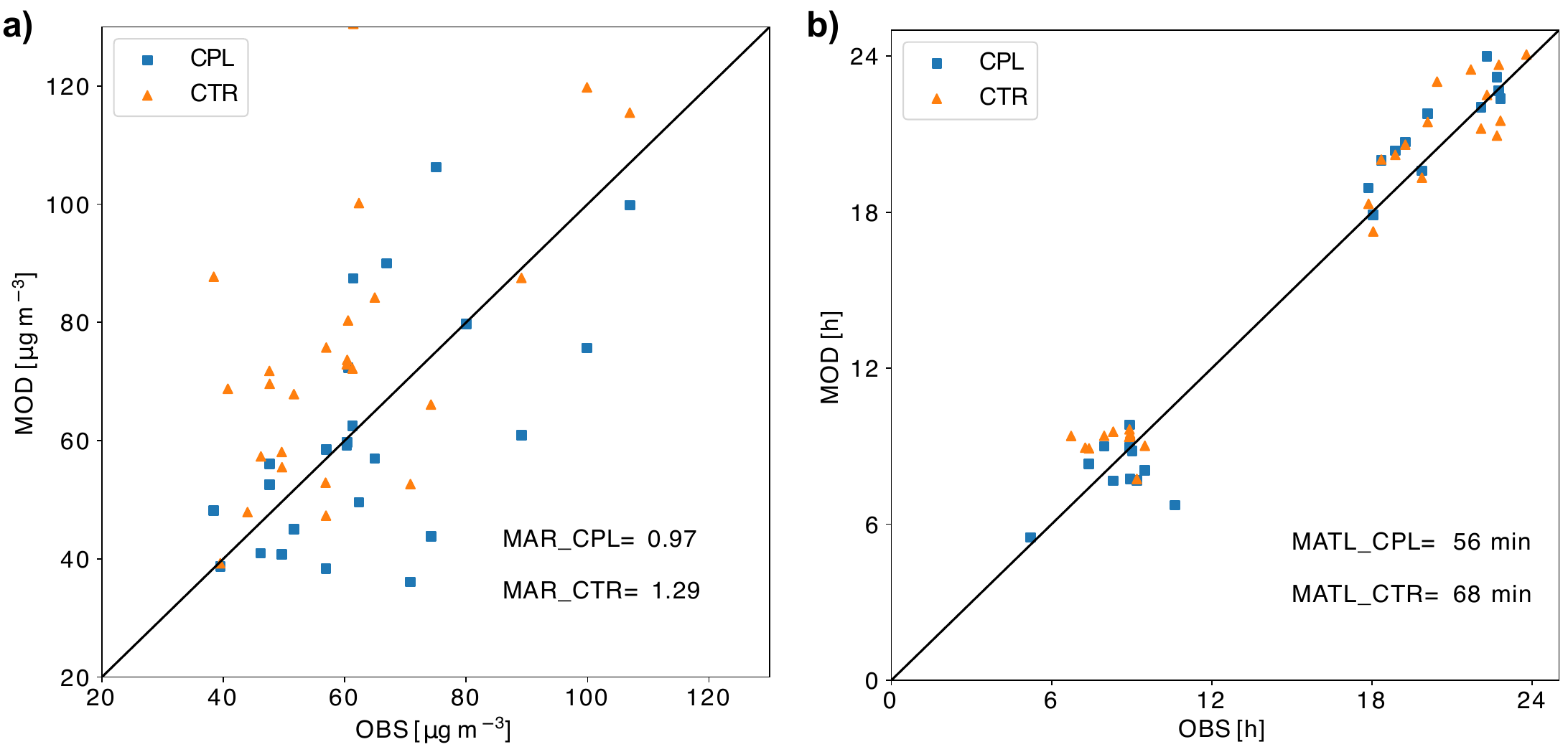}
\end{center}
\caption{Diagnostic scatter plots of matched observation--model peak magnitudes (a) and peak timings (b) for all simulations. The inset values summarize the mean amplitude ratio (MAR) in panel (a) and the mean absolute time lag (MATL) in panel (b) for CPL and CTR.}
\label{fig:peak_timings}
\end{figure}

To assess the representation of exposure-relevant short-term events, a dedicated peak analysis was performed based on the observed and modelled NO$_2$ time series, following the procedure described in Section \ref{sec:eval_metrics}. Table~\ref{tab3} summarizes the resulting peak agreement metrics, while Figure~\ref{fig:peak_timings} provides a more differentiated view of matched peak magnitudes and timings.\\

In total, 30 peaks were identified in the observation series. Relative to these peaks, CPL shows a slightly higher number of missed events than CTR (6 vs.\ 4), but it produces far fewer false peaks (2 vs.\ 8). The false alarms in CTR occur mainly during periods, especially weekends, when the static temporal emission profiles appear unable to represent the observed reduction in traffic-related NO$_2$.\\

For matched peaks, CPL reproduces peak amplitudes much more accurately overall. Across all evaluated series, the mean amplitude ratio is 0.97 for CPL, indicating an almost unbiased representation of peak magnitude, whereas CTR yields a mean amplitude ratio of 1.29 and thus a clear overall overestimation. This contrast is particularly evident at LL during T1, where CPL remains close to the observations (MAR = 1.06), while CTR overpredicts peak amplitudes strongly (MAR = 1.64). At LC during T1, CPL underestimates peak amplitudes (MAR = 0.75), which is consistent with the general negative bias already visible in the hourly NO$_2$ time series.\\

Differences in peak timing are smaller. Both simulations reproduce the timing of matched peaks with mean absolute lags on the order of one hour, with CPL showing a slight advantage overall (57\,min vs.\ 68\,min). Thus, the main benefit of the dynamic emission representation lies less in a substantial shift of peak timing skill than in reducing false hotspot peaks and improving the magnitude of simulated NO$_2$ peaks.\\

\begin{table}[h]
\caption{Peak agreement metrics for the NO$_2$ observation and model time series shown in Figure~\ref{fig:no2_time_series}. Here, \#p denotes the number of observed peaks, \#miss the number of missed peaks, \#false the number of false peaks, MAR the mean amplitude ratio, and MATL the mean absolute time lag.}
\label{tab3}
\begin{tabular*}{\textwidth}{@{\extracolsep\fill}lccccccccc}
\toprule
& & \multicolumn{4}{@{}c@{}}{CPL} & \multicolumn{4}{@{}c@{}}{CTR} \\\cmidrule{3-6}\cmidrule{7-10}
Time series & \#p & \#miss & \#false & MAR & MATL\footnotemark[2] & \#miss & \#false & MAR & MATL\footnotemark[2]\\
\midrule
LC-T1 & 9  & 3 & \textbf{2} & 0.75 & \textbf{53} & \textbf{1} & 4 & \textbf{0.99} & 88\\
LL-T1 & 10 & \textbf{1} & \textbf{0} & \textbf{1.06} & 69 & 2 & 2 & 1.64 & \textbf{59}\\
LC-T2 & 11 & 2 & \textbf{0} & \textbf{1.03} & \textbf{48} & \textbf{1} & 2 & 1.26 & 62\\
\midrule
Total & 30 & 6 & \textbf{2} & \textbf{0.97} & \textbf{57} & \textbf{4} & 8 & 1.29 & 68\\
\botrule
\end{tabular*}
\footnotetext[2]{in units of \SI{}{\minute}}
\end{table}

\section{Discussion}\label{sec:discussion}

\subsection{Scope and limitations of the comparison}\label{sec:disc_claim}

The comparison between CTR and CPL provides direct evidence that a  detector-informed traffic-emission representation can improve hyperlocal NO$_2$ modelling beyond a static inventory baseline. The strongest evidence lies not only in the global metrics but in the pattern of improvements: fewer false peaks, more realistic peak amplitudes, and clearer gains at the street-canyon site. This supports the interpretation that dynamic traffic information matters most where hyperlocal exposure is shaped by rapidly varying local activity.\\

At the same time, the study does not imply that traffic-emission realism is the single dominant control on all observed NO$_2$ variability. The persistence of substantial error at LC during T1 shows that remaining uncertainty in other emission sectors, receptor representativeness, boundary conditions, chemistry, or local meteorology can outweigh the benefits of better traffic emissions at individual sites. In this sense, the present comparison should be interpreted as an added-value experiment under realistic but still imperfect urban-model conditions, not as proof that traffic realism alone is sufficient to resolve all hotspot mismatches.

\subsection{Why dynamic emissions provide more reliable estimates}\label{sec:disc_reliability}

Dynamic emissions scale locally with calibrated traffic flows, while static emission maps still rely on highly approximate disaggregation approaches of national totals to finer spatial resolutions. Such downscaling procedures often fail to capture local heterogeneity in traffic intensity, fleet composition, and driving conditions, which can lead to systematic spatial biases \citep{Tsanakas2017}. This, among other factors, offers an explanation for the observed reduced local bias and apparent more consistent model performance of the dynamic approach CPL compared to CTR across different sites.\\

Furthermore, the often limited timeliness of static emission inventories represents a critical limitation. These datasets are typically compiled and used with considerable time lags and may therefore not reflect recent developments in vehicle technology, fuel standards, or regulatory measures. As a result, their application can introduce additional uncertainty if not treated with caution -- especially in contexts where rapid technological advancements and policy interventions (e.g., the implementation of low-emission zones or fleet electrification strategies) have already led to substantial emission reductions. This temporal mismatch can be particularly problematic in urban areas undergoing fast transitions in mobility systems.\\

In this regard, microscopic traffic models provide greater flexibility and adaptability as they allow for relatively straightforward updates of fleet composition, emission factors, and traffic demand, enabling a more timely representation of current conditions. This combination of temporal responsiveness and process-level detail makes dynamic approaches particularly well suited for high-resolution air quality modeling and policy assessment, compared to the more rigid methodologies underlying static emission inventories.

\subsection{Why the improvement is site dependent}\label{sec:disc_site}

The contrast between LL and LC is scientifically informative. LL is a classic street-canyon hotspot in which reduced ventilation amplifies the hyperlocal traffic emission impact. Under such conditions, biases introduced by smoothed temporal profiles and static emission representations are more likely to propagate into simulated concentrations, leading to the observed overestimation of peak values and the emergence of spurious peaks. In contrast, the dynamic SUMO-based representation responds more realistically to changing traffic intensity and traffic state. LC, in contrast, is more open and thus more exposed to background variability, nearby non-traffic contributions, and the precise spatial representativeness of the receptor point. A better road-traffic estimate can therefore yield only limited gains if one of these additional terms remains biased.\\

This evidence of site dependence is consistent with the broader street-canyon literature, which shows that geometry, ventilation, and chemistry strongly modulate how source signals are translated into local concentration hotspots \citep{Vardoulakis2003,Zhong2016,Dai2022}.

\subsection{Remaining uncertainty budget}\label{sec:disc_uncertainty}

Three uncertainty groups remain particularly relevant. First, traffic uncertainty is reduced but not eliminated in CPL because detector coverage is incomplete and the calibration cannot constrain all roads equally well. This limitation is especially relevant near LC, where missing detector information likely reduced the quality of local traffic reconstruction. More generally, the available detector network is not dense enough to calibrate all roads in the city with comparable confidence.\\

Second, the non-traffic emission background remains static in both experiments, meaning that mismatches in residential, industrial, or other mobile sectors can still contaminate the interpretation of traffic-driven NO$_2$ at the receptors. This is particularly important for LC during T1, where the negative bias in CPL suggests that additional local sources or underrepresented non-traffic emissions may have contributed to the observed concentrations. In that case, the apparent advantage of CTR in some bulk metrics should not be interpreted as evidence for a superior traffic-emission representation, but rather as a possible compensation of errors across source sectors.\\

Third, chemistry and transport uncertainty persists because even a physically rich LES framework relies on uncertain mesoscale meteorological and air composition boundary conditions, as well as simplified chemistry. Meteorological factors that affect the diurnal evolution of boundary-layer mixing, local dispersion pathways and NO$_2$ chemistry may further modulate the observed concentration peaks and can therefore partly mask the effect of improved traffic emissions.\\

\subsection{Implications for hyperlocal air-quality modelling and urban digital twins}\label{sec:disc_implications}

The main implication of this work is methodological and practical: for hyperlocal NO$_2$ applications, it is more informative to ask \emph{where} dynamic traffic emissions improve the model and \emph{which performance dimensions} improve than to ask whether they improve all metrics everywhere. Dynamic emissions add the most value for hotspot realism, for the suppression of false peaks, and for the representation of peak amplitude at traffic-dominated receptors.\\

That finding is relevant for operational urban digital twins, for policy-oriented studies that target short-term exposure, adaptive traffic interventions, or hotspot mitigation, and for urban planning processes that increasingly rely on spatially explicit environmental evidence. More specifically, improved hotspot realism at the street and intersection scale increases the credibility of urban digital-twin applications and planning-support contexts that rely on scenario comparison, forward-looking traffic assessment, or the evaluation of targeted interventions before implementation. This includes not only hotspot-oriented mitigation in the narrow sense, but also the air-quality appraisal of future traffic-management options such as rerouting strategies, adaptive signal control, or context-specific information measures, as well as the environmental assessment of urban design and planning decisions. It also provides a transparent benchmark against which future extensions can be evaluated, including more complete non-traffic dynamics, improved traffic sensing, richer chemistry, or explicit non-exhaust emissions.\\

At the same time, the present study provides a realistic demonstration of what an urban air-quality digital twin can achieve under currently available data conditions rather than under idealized experimental conditions. This makes the results relevant not only as a methodological benchmark, but also as a practical reference for cities in which real-time traffic information remains incomplete and unevenly distributed.\\

Overall, the present comparison supports a focused conclusion: replacing static road-traffic emissions with detector-informed dynamic emissions provides measurable added value for hyperlocal NO$_2$ modelling, especially when the target is hotspot realism and peak representation rather than only city-wide mean fields. The clearest gains are found at the street-canyon hotspot, where locally evolving traffic states interact most strongly with reduced ventilation, and in the improved representation of peak amplitudes and false peaks. Although the present paper demonstrates these effects using selected hotspot-scale examples, the underlying modelling framework is not limited to such local application windows, but is in principle capable of resolving NO$_2$ fields at comparable spatial detail across the full urban domain. At the same time, the remaining site-specific mismatches show that traffic realism should be treated as one major component of the urban NO$_2$ uncertainty budget, but not as its sole solution. The present results therefore matter not only for retrospective hotspot analysis, but also for future urban workflows in which traffic and environmental models are used anticipatively to compare options, identify likely air-quality impacts, and support more environmentally sensitive mobility management and urban planning processes.


\section{Methods}\label{sec:methods}

\subsection{Study design and hypotheses}\label{sec:design}

The study follows a controlled two-experiment design. The reference experiment CTR represents all emission sectors, including road traffic, with static inventories and prescribed temporal factors. The coupled experiment CPL replaces the static road-traffic sector inside the SUMO domain by dynamic emissions derived from mesoscopic traffic simulations, while keeping all other sectors identical to CTR. Meteorological forcing, boundary conditions, chemistry, numerics, and receptor extraction are otherwise unchanged between the two experiments. This design allows differences in simulated NO$_2$ to be attributed as directly as possible to the representation of road-traffic emissions.\\

The primary hypothesis is that CPL yields more robust hyperlocal NO$_2$ performance than CTR when evaluated against hourly observations. A secondary hypothesis is that this added value is strongest at locations where traffic dynamics and reduced ventilation interact most strongly, i.e. at the street-canyon site. A third hypothesis is that the largest gains occur for peak amplitude and false-peak suppression, whereas peak timing may remain partly controlled by meteorology and background chemistry.

\subsection{Study area, validation periods, and observational basis}\label{sec:studyarea}

The application domain is the city of Leipzig, Germany, with evaluation focused on two official traffic-oriented monitoring locations: Leipzig Center (LC), situated in a relatively open intersection environment near the main station, and Leipzig L\"utznerstra\ss e (LL), situated within a narrow street canyon in the western part of the city. More specifically, LC is located roughly \SI{300}{m} west of Leipzig Main Station at the intersection of Willi-Brandt-Platz and Am Hallischen Tor, with the nearest buildings located about \SI{30}{m} to the south. By contrast, LL is located in the western district of Leipzig Lindenau within a narrow street canyon of approximately \SI{15}{m} width. The contrasting morphology of these sites is central to the study design because it allows us to test whether the benefit of dynamic emissions depends on hotspot type \citep{weger2022}.\\

\begin{figure}[H]
\vspace*{2mm}
\begin{center}
\includegraphics[width=1.0\textwidth]{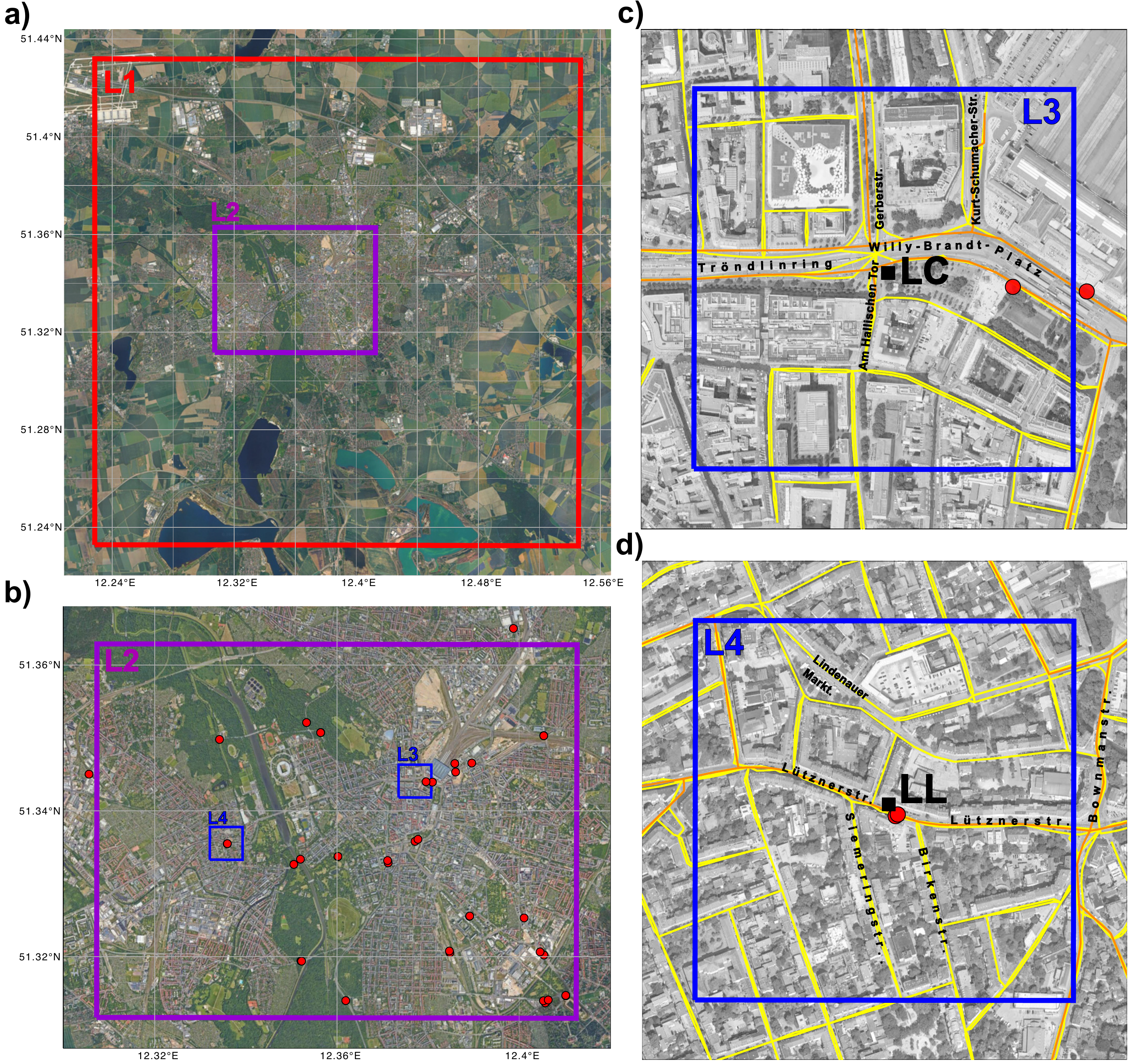}
\end{center}
\caption{CAIRDIO simulation setup for Leipzig consisting of domains L1 (\SI{60}{m}), L2 (\SI{20}{m}), L3 (\SI{5}{m}), and L4 (\SI{5}{m}). The main study areas are identified by domains L3 and L4, both spanning $1\times1$~km$^2$ and centred on the official air monitoring sites Leipzig Center (LC) and Leipzig L\"utznerstra\ss e (LL), respectively. For these areas, the SUMO traffic network is shown by yellow lines, while the line emissions of the static inventory are shown in orange. Traffic observation sites used for calibration in SUMO are marked by red filled circles within domains L2, L3, and L4. Background maps are based on Google Maps satellite imagery (© Google)}
\label{fig:domains}
\end{figure}

The CAIRDIO and SUMO setup is illustrated in Figure~\ref{fig:domains}. The city-scale domain is represented by L1 at \SI{60}{m} horizontal resolution, nested by L2 at \SI{20}{m} resolution over the central urban area. The two innermost evaluation domains L3 and L4 provide \SI{5}{m} resolution around LC and LL, respectively. These inner domains define the two principal hotspot environments analyzed in the present study.\\

The selection of suitable validation periods was based on a screening of the year 2025 with respect to traffic-detector availability and meteorological suitability. Periods dominated by strong synoptic advection, high wind speeds, or precipitation were deprioritised because local traffic emissions are then less decisive for near-road concentrations. Since traffic-flow observations at calibration sites are typically used at \SI{5}{min} intervals, daily detector reporting rates were used as one key criterion for period selection. In SUMO, multi-lane roads are represented by single directed edges, implying that all lane-specific calibrators assigned to one edge must report simultaneously for an observation to be usable. Daily reporting rates were therefore computed as the ratio between valid reports and the maximum possible number of 288\,=\,1440\,[min/day] / 5\,[min] values per day.\\

\begin{figure}[H]
\vspace*{2mm}
\begin{center}
\includegraphics[width=1.0\textwidth]{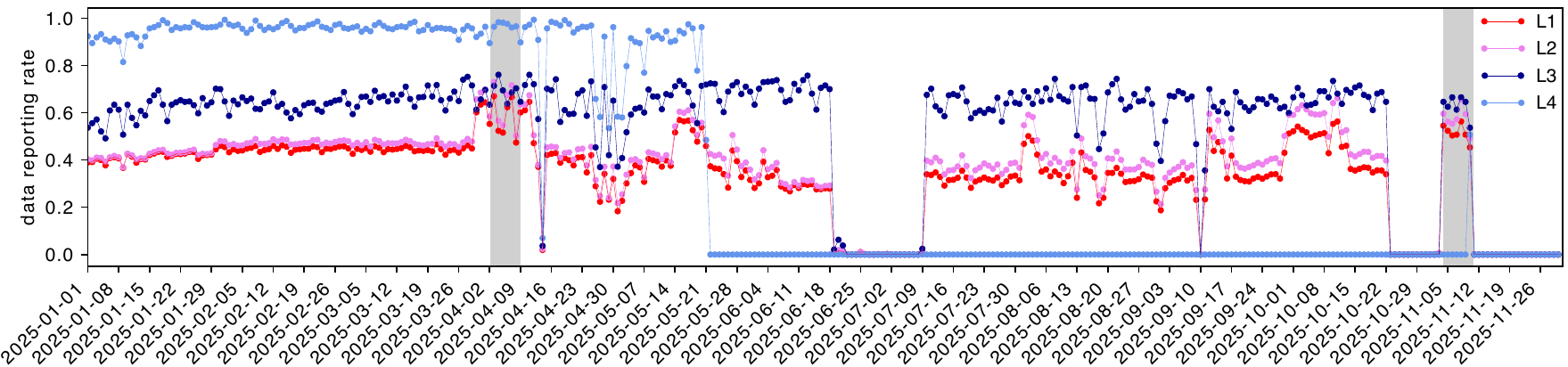}
\end{center}
\caption{Daily average traffic detector reporting rates for domains L1, L2, L3, and L4, respectively. The shaded windows indicate the two validation periods selected for the present study.}
\label{fig:det_report_rate}
\end{figure}

Figure~\ref{fig:det_report_rate} shows the resulting detector reporting rates for domains L1--L4. Several periods with above-average detector availability are apparent. For the present analysis, two one-week periods were selected: T1 from 2~April 00:00~UTC to 9~April 00:00~UTC, and T2 from 4~November 00:00~UTC to 10~November 00:00~UTC. During T1, mean reporting rates reached 0.58 (L1), 0.62 (L2), 0.69 (L3), and 0.95 (L4), all above their respective annual median values. Meteorological conditions during this period were mostly stable, with high pressure and low cloud cover prevailing on most days; only 5~April showed temporarily stronger north-easterly winds associated with a cold-air outbreak, but without relevant precipitation in Leipzig. During T2, mean reporting rates remained above the annual median in L1--L3, with values of 0.45, 0.50, and 0.55, respectively. This period was characterised by calm conditions, frequent inversions, and persistent stratocumulus or high-fog conditions during the later days. Because detector coverage in domain L4 was inactive during the second half of 2025, site LL is only evaluated for T1.

The observational basis for NO$_2$ validation is provided by the official monitoring network operated by the S\"achsisches Landesamt f\"ur Umwelt, Landwirtschaft und Geologie (LfULG, Dresden, Germany). At both sites, NO$_2$ is measured by the chemiluminescence reference method \citep{Fontijn1970}, which is prescribed by European air-quality legislation for continuous NO and NO$_2$ monitoring. The stations therefore provide an appropriate benchmark for evaluating the ability of the model chain to reproduce traffic-driven NO$_2$ dynamics under regulatory monitoring conditions. The associated quality-assurance framework includes regular instrument maintenance and periodic review by national reference laboratories, with a target measurement uncertainty of no greater than 15\% for NO$_2$.\\

\subsection{Traffic model and detector assimilation in SUMO}\label{sec:sumo}

Traffic dynamics were simulated with SUMO, using the Leipzig  operational network and the detector-informed calibration workflow developed for the urban digital-twin setting \citep{Floetteroed_2022,Keler2023}. The basic infrastructure for SUMO is a routable road network for different modes of transport covering pedestrians, bicycles, cars, trams, buses and lorries. This static network is primarily derived from OpenStreetMap (OSM) and supplemented with additional local datasets. At the signalized intersections on main roads the records of signal plans are implemented where available.\\

Time-resolved information about population activity and commuters, aggregated in zones of interest for living, working, shopping, and leisure is required for estimating the base travel demand i.e. the route sets representing people movements. Therefore, an existing MATSim demand model \citep{matsim-leipzig} was used and refined using the traffic volume map for 2023 provided by the City of Leipzig. This setup represents a typical traffic situation and serves as a baseline for further scenario simulations.\\

As SUMO is applied to the comparatively large-scale road network of Leipzig, computational constraints necessitate the use of its mesoscopic simulation mode \citep{sumo_meso_doc}. In this mode, traffic is still represented at the level of individual vehicles, as in the microscopic formulation. However, vehicle movements are derived from a queue-based model \citep{diss_eissfeldt} that incorporates mesoscopic flow properties (e.g., current density and lane capacity) and are handled in discrete units of space and time within an event-based framework. This approach replaces computationally more demanding car-following models, which explicitly simulate gap acceptance and provide quasi-continuous updates.\\

In the digital-twin framework, link-level traffic flows and speeds are further constrained by stationary detector information where available. The applied online calibration process consists of two parts, i.e. data processing and simulation with calibration and prediction. The whole process chain is implemented, customized and enhanced for the simulation setup for Leipzig city. The data process part includes data correction, data aggregation, data fusion and data extrapolation. Currently, traffic measurements (speed and flow) from stationary sensors, provided by the city of Leipzig via Mobilithek are considered in the data processing, where flow types include passenger cars and trucks. Speed average at each detector group is weighted with respect to the number of flows, and the interval-based data quality indicator is calculated according to the number of available data records in a pre-defined interval. If two or more detector groups locate at the same edge, average values are derived and used. The observed traffic demand per pre-defined interval for calibration and the route sets from the base traffic demand are then used together with the given network as inputs in the simulation. \\

\begin{figure}[H]
\vspace*{2mm}
\begin{center}
\includegraphics[width=1.0\textwidth]{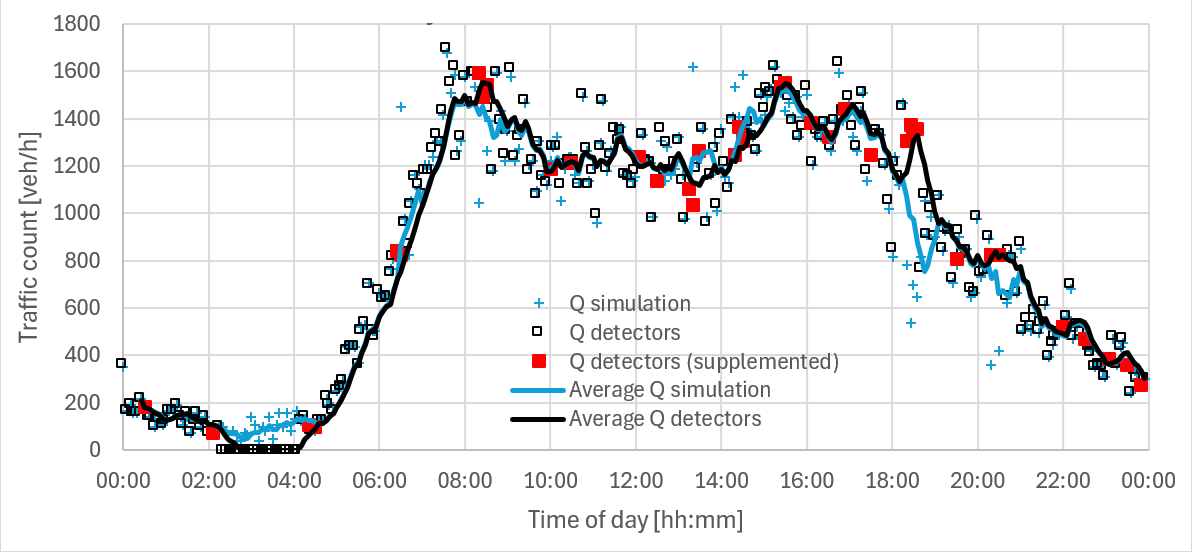}
\end{center}
\caption{Comparison of detector (black square markers) and SUMO simulation data (blue crosses) for traffic flow at Leipzig main railway station in eastern direction on 17 June 2024. Supplemented data points are highlighted in red.}
\label{fig:calibration_data_supp}
\end{figure}

To ensure consistent simulation inputs despite data uncertainties, automated data supplementation routines are applied: 
\begin{enumerate}
\item In case of no data available for a detector for one day, the detector may be “ignored” for this period, or data from a comparable day (e.g. the same weekday in a previous week) may be used instead. \item In case of data missing in short time periods for a few minutes, the data can be estimated based on the data already received up to the current time of this day, which is shown as an example in Figure~\ref{fig:calibration_data_supp}. In average for the data received for Leipzig so far about 10\% of the data are supplemented.
\end{enumerate}

The local calibration performance relevant for the present study is illustrated in Figure~\ref{fig:sumo_vs_det_time_series}, which compares modelled and observed traffic flows and speeds at the detector pairs in the vicinity of LC and LL during the selected validation periods. In the context of this manuscript, SUMO is not treated as the endpoint of the analysis, but as the source of dynamic and physically interpretable traffic-activity information for downstream emission generation. The key methodological role of the traffic model is therefore to provide a temporally resolved and observation-constrained representation of local traffic states that can be translated into dynamic road-traffic emissions.

\subsection{CAIRDIO setup, chemistry, and nesting}\label{sec:cairdio}

We use the City-scale AIR dispersion model with DIffuse Obstacles (CAIRDIO) to simulate urban dispersion and chemical transformation of traffic-related NO$_x$. CAIRDIO is based on a large-eddy simulation (LES) framework that resolves turbulent interactions between the atmospheric boundary layer, buildings, vegetation, terrain, and thermally driven surface heterogeneity \citep{weger2021,weger2022}. In addition to momentum transport, the model represents radiative interactions as well as heat and moisture budgets of surface elements, including building facets and vegetation layers, which is important for reproducing urban thermal stratification under varying meteorological conditions. This makes CAIRDIO suitable for analysing hyperlocal concentration variability in environments where street-scale ventilation and recirculation strongly affect pollutant residence times \citep{Vardoulakis2003,Zhong2016}.\\

The Leipzig setup consists of two main nested CAIRDIO domains and two innermost high-resolution evaluation domains (Figure~\ref{fig:domains}). The outer domain L1 covers city-scale meteorology and air quality using a \SI{60}{m} horizontal grid spacing, while L2 refines the central urban area to \SI{20}{m}. The 1$\times$\SI{1}{\kilo\metre\squared} evaluation domains L3 and L4 have \SI{5}{m} horizontal grid spacing, which is needed for resolving the short source-receptor distances at the two traffic-oriented sites LC and LL, respectively. Vertically, domain L1 extends up to a height of $1\,\unit{km}$ for and adequate capture of boundary layer dynamics even under convective conditions. The nested domains do not extent beyond heights of \SI{300}{m} (L2) and \SI{100}{m} (L3 and L4) above ground, as interpolated boundary conditions from respective parent domains are prescribed at all lateral and top-domain boundaries with a high temporal frequency matching the parent-domain integration time step. Near-surface vertical grid spacing is \SI{7}{m} in L1, \SI{5}{m} in L2, and \SI{3}{m} in L3 and L4. \\

To drive the model, external boundary conditions for L1 are derived from precursor simulations with the ICON numerical weather prediction model for meteorology \citep{zaengl2015,zaengl2022} and from ECMWF-CAMS analysis data for atmospheric composition \citep{camsdata}. The ICON precursor simulations are themselves driven by ICON-EU analysis data provided by the German Weather Service (DWD) \citep{iconeu}. Further details on the representation of urban landscape, topography, and the associated static input data in CAIRDIO are given by \citet{weger2022} and the follow-up model documentation.\\

Chemical transformation is represented by the Generic Reaction Set (GRS) \citep{azzi1992}, which includes 7 key reactions representing the fast NO--NO$_2$--O$_3$ system and the influence of reactive organic compounds in a simplified form on it. In addition to transport and chemistry, CAIRDIO also represents dry and wet deposition. The present chemistry setup is appropriate for the attribution experiment because it captures the short-timescale processes most relevant to near-road NO$_2$ without introducing a substantially larger computation overhead. At the same time, this simplification should be kept in mind when interpreting site-specific residual bias, especially in street-canyon environments \citep{Zhong2016,Dai2022}.\\

\subsection{Static and dynamic emission representation}\label{sec:emissions}

Emission representation in CAIRDIO follows the Selected Nomenclature for Air Pollution (SNAP) sector framework \citep{eea2003}. In the static reference setup, all sectors are represented by annual-mean emission maps for Germany provided by the Federal Environmental Agency (UBA) except for the traffic emissions (SNAP 7) within the city region, which were provided as spatially more accurate line sources by the municipality of Leipzig \citep{weger2022}. This allowed a direct rasterisation onto the simulation grids without geometric smoothing losses by geometric line--grid intersection in the Lambert Azimuthal Equal-Area projection. Since, however, the UBA-based emission maps with an original resolution of $0.5\times0.5$~km$^2$ are insufficient for microscale applications, a GIS-based disaggregation workflow was used to project the emissions to the CAIRDIO grids at \SI{60}{m}, \SI{20}{m}, and \SI{5}{m} resolution. Further details of this aspect are given in \citet{weger2022}.\\

\begin{figure}[H]
\vspace*{2mm}
\begin{center}
\includegraphics[width=0.6\textwidth]{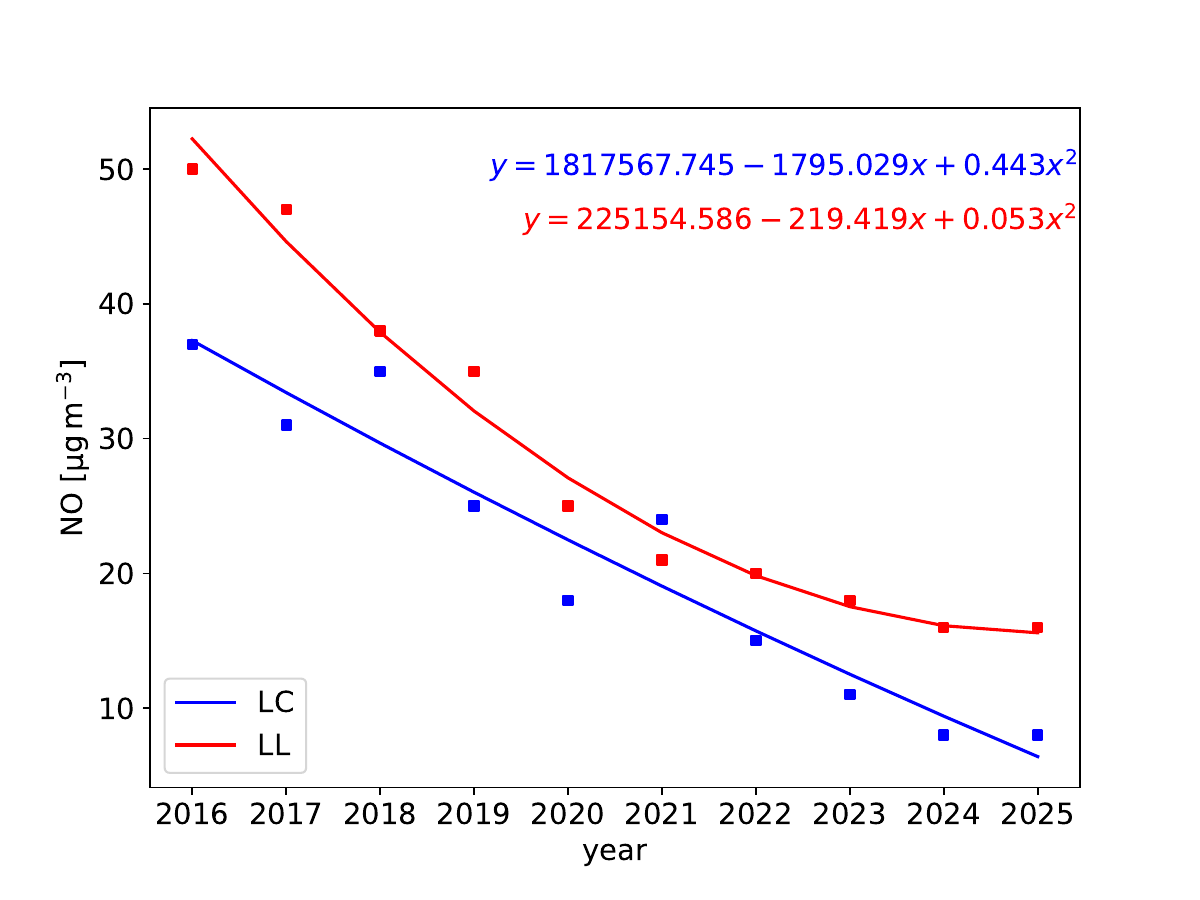}
\end{center}
\caption{Observed annual mean NO concentration trends at sites LC (blue markers) and LL (red markers) over the period 2016 to 2025, which serve as a proxy for traffic NO$_x$ emission reductions. The blue and red lines with equations show respective quadratic least-squares fits to the data.}
\label{fig:NO_annual_mean_trends}
\end{figure}

As the used emission datasets refer to the year 2016 they do not represent recent vehicle emission standards in 2025. To account for traffic NO$_x$ reductions over the stated period, a reduction factor of 0.3 is applied to the SNAP 7 emissions. This value is a conservative estimate and justified by annual mean NO trends at the traffic-oriented sites LC and LL as shown in Figure~\ref{fig:NO_annual_mean_trends}. Accordingly, NO concentrations declined by $70\,\%$ at site LC and by $83\,\%$ at site LL over the period 2016 to 2025. \\

To derive time-specific emissions from the static inventories, monthly, weekly, and hourly sector-specific temporal factors were applied as specified in \citet{tno2011}. In the CTR setup, this static procedure is used for all sectors, including road traffic. The resulting traffic-emission baseline therefore corresponds to a conventional ``static + standard time-profile'' representation.\\

In the dynamic setup, SUMO is used to simulate vehicular pollutant emissions based on the database application HBEFA in version 4.2.. For each vehicle the emissions are calculated based on its dynamic movement and summed up link-based. The link-based emissions are distributed by their share of length to the CAIRDIO grids. All non-road sectors remain identical to the static setup, and static traffic emissions are retained outside the SUMO domain. This design yields a clean contrast between a conventional static road-traffic representation and a detector-informed dynamic one. Compared with recent data-driven inventory developments, the present workflow belongs to the broader class of activity-aware emission modelling, but operates at finer urban-process resolution through direct coupling with an on-line calibrated traffic model \citep{wu2020,DRIVE2025}.\\

\subsection{Experiment definition and receptor extraction}\label{sec:experiments}

For each validation period, two experiments are performed. In the reference experiment CTR, all emission sectors are represented by the static emission workflow. In the coupled experiment CPL, static road-traffic emissions are replaced inside the SUMO domain by SUMO-derived dynamic traffic emissions, while traffic outside the SUMO domain and all remaining sectors are kept identical to CTR. This setup defines a targeted sensitivity experiment in which the representation of road-traffic emissions is the only intended difference between the two simulations.\\

To ensure that differences in simulated NO$_2$ can be attributed as directly as possible to the traffic-emission representation, both experiments use identical meteorological and chemical drivers. In particular, the same precursor-generated meteorological fields, including wind speed, subgrid-scale diffusion coefficients, and deposition rate coefficients, are applied in CTR and CPL. The actual dispersion simulation then computes emissions, three-dimensional advection, turbulent diffusion, deposition, and chemical transformation for the species included in the GRS mechanism, namely NO, NO$_2$, O$_3$, and reactive organic compounds (ROC).\\

Modelled concentrations are extracted at model run-time from the innermost \SI{5}{m} evaluation domains at the first vertical model level using nearest-neighbour sampling at the monitoring-site locations. In a post-processing step, the high-frequency time series are hourly averaged to adapt them to the observation space. The subsequent analysis is based on paired hourly observation--model time series for the three evaluated cases LC-T1, LL-T1, and LC-T2.\\

\subsection{Evaluation metrics and peak analysis}\label{sec:eval_metrics}

Model performance is evaluated on the basis of paired hourly observed and simulated NO$_2$ concentration series for LC-T1, LL-T1, and LC-T2. General model performance is assessed with fractional bias (FB), centred root mean square error (CRMSE), coefficient of determination ($\mathrm{R}^2$), and Willmott's index of agreement (IOA) \citep{Willmott1981}. These metrics are reported for each individual time series and across all time series combined. They are complemented by a targeted peak analysis because the main claim of the paper concerns the representation of hotspot-relevant concentration peaks rather than only average agreement.\\

Peaks are identified in both observed and modelled NO$_2$ time series using Python-based \texttt{scipy.signal.find\_peaks} with the parameterisation adopted to the given data characteristics ($\mathrm{wlen}=24$, $\mathrm{prominence}=10$, $\mathrm{distance}=6$). An additional absolute threshold of \SI{20}{\micro\gram\per\cubic\meter} is further applied to exclude non-significant events. Observation and model peaks are then matched within a symmetric \SI{6}{h} time window. Unmatched observed peaks are counted as misses, whereas unmatched model peaks are counted as false alarms. Because the number of misses and false alarms differs between CTR and CPL, the matched peak samples may also differ between the two experiments.\\

For matched events, peak magnitude and peak timing are derived from a moment-based peak characterization rather than from the absolute peak maximum alone. After identifying peak bases, the zeroth, first, and second peak moments are computed, and peak timing and magnitude are then derived using a centre-of-gravity approach. This procedure reduces sensitivity to peak asymmetry and to multiple nearby local maxima within the same broader event. Final peak performance is summarized through the mean amplitude ratio (MAR) and the mean absolute time lag (MATL) between observed and modelled peak series.\\

\backmatter

\section*{Data availability}

CAIRDIO and SUMO modelling results, traffic detector observations, air quality observation, and further supplementary files and scripts to reproduce all Figures and Tables in the manuscript are accessible at \url{https://doi.org/10.5281/zenodo.19135030}.

\section*{Code availability}
The source code of the CAIRDIO model is accessible in release under the license GPL v3 and later at \url{https://doi.org/10.5281/zenodo.6075354}.
The source code of SUMO is licensed under EPL 2.0. Version 1.26.0 was used (\url{https://doi.org/10.5281/zenodo.18406080}). SUMO documentation can be found at  \url{https://sumo.dlr.de/docs/index.html}.

\section*{Acknowledgements}
The language of the manuscript was partially improved using LLMs, however, no content generated by these tools was incorporated without careful review and verification. 
Any satellite images used in the Figures of the manuscript are from Google Maps (© Google).
German-wide emission data were provided by the German Environment Agency (Umweltbundesamt, UBA). Line emissions for road transport in Leipzig were provided by the Saxon State Office for Environment, Agriculture and Geology (Sächsisches Landesamt für Umwelt, Landwirtschaft und Geologie, LfULG), Environment and Transport Information System (Fachinformationssystem Umwelt und Verkehr, FIS UUV). Measurements of PM10 and other meteorological variables from the consulted air-monitoring sites were provided by the LfULG. Meteorological ICON-EU analysis fields used to drive the CAIRDIO simulations were downloaded from the Pamore platform of the Deutscher Wetterdienst (DWD). CAMS European Air Quality forecast data used to drive the CAIRDIO simulations were retrieved from the ECMWF Atmosphere Data Store. Traffic detector data was provided by the the City of Leipzig via Mobilithek (Dynamische Detektordaten Stadt Leipzig. \url{https://mobilithek.info/offers/110000000002877002}, Statische Detektordaten Stadt Leipzig. \url{https://mobilithek.info/offers/110000000002877001})

This work contributes to the activities of the project CoKliP - Cockpit Klimamission und Planungsbeschleunigung, funded by the German Federal Ministry of Research, Technology and Space (BMFTR) under grant numbers 01UR2520B (subproject B: Integration klimarelevanter Umwelt-, Geo- und Fernerkundungsdaten in kommunale Planungsprozesse) and 01UR2520D (subproject D:  Luftqualitäts- und Strömungssimulationen, KI-gestützte beschleunigte Modellierung
von Luftqualitäts- und Klimadaten).

\section*{Funding}
The research is part of the project Artificial Intelligence and Mobility (AIAMO), funded by the German Federal Ministry for Digital and Transport (BMDV) under Founding 45KI19D041 (BMVI BAV).\\

\section*{Author contributions}
Michael Weger: Conceptualization, Data curation, Investigation, Methodology, Software, Validation, Visualization, Writing -- original draft preparation. Thomas Trabert: Conceptualization, Investigation, Supervision.  Timo Houben: Conceptualization, Investigation, Supervision. Alexander Sohr: Software, Data curation, Investigation, Writing -- original draft preparation.  Elmar Brockfeld: Software, Data curation, Investigation, Writing -- original draft preparation. Oswald Knoth: Conceptualization, Investigation, Supervision.  Roland Schrödner: Conceptualization, Investigation, Supervision. Jan Bumberger: Conceptualization, Investigation, Supervision, Writing -- original draft preparation, Funding acquisition, Resources, Project administration.  

\section*{Competing interests}
The authors declare no competing interests.





\bibliography{260315_no2-hotspots_v2.bib}

\end{document}